\documentclass[11pt,aps,prd,nofootinbib,superscriptaddress,preprint,preprintnumbers
]{revtex4-2}

\usepackage{amssymb}
\usepackage{amsmath}
\usepackage{bbm}
\usepackage{graphicx}
\usepackage{hyperref}
\usepackage{multirow}
\usepackage{booktabs}
\usepackage{array}
\usepackage{xcolor}
\usepackage{rotating}

\usepackage{physics} 
\usepackage{slashed} 
\usepackage{textcomp}

\usepackage{makecell}

\allowdisplaybreaks[2]

\newcommand{\SKLP}{State Key Laboratory of Particle Detection and Electronics, University of Science and Technology of China, Hefei 230026, Anhui, People’s Republic of China}
\newcommand{\USTC}{Department of Modern Physics, University of Science and Technology of China, Hefei 230026, Anhui, People’s Republic of China}
\newcommand{\ACFS}{Anhui Center for Fundamental Sciences in Theoretical Physics, University of Science and Technology of China, Hefei 230026, Anhui, People’s Republic of China}
\newcommand{\NUS}{Department of Physics, National University of Singapore, Singapore 117551, Singapore}

\newcommand{\Hline}{\noalign{\hrule height 1.2pt}}
\newcommand{\depbox}[1]{\makebox[1.5em][c]{#1}} 

\begin{document}

\title{\boldmath Doubly charged Higgs production within the Higgs triplet model at future electron-positron  colliders}

\author{Shu-Xiang Li}
\affiliation{\SKLP}
\affiliation{\USTC}

\author{Ren-You Zhang}
\email{zhangry@ustc.edu.cn}
\affiliation{\SKLP}
\affiliation{\USTC}
\affiliation{\ACFS}

\author{Ming-Hui Liu}
\affiliation{\SKLP}
\affiliation{\USTC}

\author{Xiao-Feng Wang}
\affiliation{\SKLP}
\affiliation{\USTC}

\author{Zhong-Yuan Liu}
\affiliation{\SKLP}
\affiliation{\USTC}

\author{Yi Jiang}
\affiliation{\SKLP}
\affiliation{\USTC}

\author{Liang Han}
\affiliation{\SKLP}
\affiliation{\USTC}

\author{Qing-hai Wang}
\email{qhwang@nus.edu.sg}
\affiliation{\NUS}

\date{\today}
\begin{abstract}
We investigate in detail the discovery potential of the doubly charged Higgs boson at the Compact Linear Collider in $e^-e^-$, $e^-\gamma$, $\gamma\gamma$, and $e^+e^-$ collision modes, within the Higgs triplet model at two extreme benchmark points as representatives of the Yukawa-like and gauge-like regions. In the Yukawa-like region, the most promising production mechanism is the single production via $e^-e^-$ and $e^-\gamma$ collisions. Given the subsequent decay of the doubly charged Higgs into a same-sign lepton pair, CLIC can achieve statistical significance well beyond the discovery threshold, within the parameter space permitted by experimental constraints. In the gauge-like region, with the $\ell^{\pm}\ell^{\pm} + \geq 3j$ final state, CLIC exhibits robust discovery potential for the doubly charged Higgs boson, up to a mass of approximately $1.2~\mathrm{TeV}$. We also investigate the search for doubly charged Higgs at the HL-LHC. Our results demonstrate that CLIC possesses greater advantages and offers superior discovery potential for the doubly charged Higgs boson, compared to the HL-LHC.

\begin{description}
\item[keywords]
Doubly charged Higgs production, Higgs triplet model, Electron-positron colliders
\end{description}
\end{abstract}

\maketitle
\newpage

\section{Introduction}
\label{sec:1}
\par
The discovery of neutrino oscillations \cite{Super-Kamiokande:1998kpq,SNO:2002tuh,KamLAND:2002uet} marks a pivotal milestone in particle physics, providing compelling experimental evidence that neutrinos have nonzero masses. Cosmological data tightly constrain the sum of neutrino masses \cite{Planck:2018vyg}, while tritium $\beta$-decay experiments provide direct, albeit presently less sensitive, upper limits on the effective electron antineutrino mass \cite{KATRIN:2024cdt}. Taken together, these observations point to a neutrino mass scale in the sub-eV range, possibly below $\mathcal{O}(10^{-1}\, \text{eV})$. However, the standard model (SM) cannot naturally account for nonzero neutrino masses. Even with right-handed neutrinos incorporated into the SM, generating neutrino masses consistent with experimental constraints solely through Dirac mass terms requires an extraordinarily small Yukawa coupling, $Y_{\nu} \lesssim 10^{-12}$, thus posing a severe naturalness problem. This tension is effectively resolved within the seesaw framework, in which heavy Majorana neutrinos generate light neutrino masses via the dimension-five operator $LLHH/\Lambda$ \cite{Weinberg:1979sa}, thereby explaining the observed neutrino mass scale without assuming an unnaturally small Yukawa coupling. The seesaw mechanism \cite{Minkowski:1977sc,Gell-Mann:1979vob,yanagida1979} can be classified into three distinct types, according to the gauge representations of the heavy fields introduced to generate the dimension-five Weinberg operator \cite{Ma:1998dn}: Type-I, introducing right-handed fermion singlets \cite{Mohapatra:1979ia}; Type-II, involving left-handed scalar triplets \cite{magg1980,Schechter:1980gr,Cheng:1980qt}; and Type-III, incorporating fermion triplets \cite{Foot:1988aq}.

\par
The Higgs triplet model (HTM) \cite{Gunion:1989ci,Hambye:2000ui,Arhrib:2011uy} provides a minimal realization of the Type-II seesaw mechanism by extending the SM scalar sector with an ${SU(2)}_L$ triplet. After electroweak symmetry breaking (EWSB), the Higgs triplet $\Delta$ acquires a vacuum expectation value (VEV) $v_{\Delta}$, inducing Majorana masses for the neutrinos through its Yukawa couplings to the left-handed lepton doublets, $m_{\nu} = \sqrt{2} v_{\Delta} Y$. The mixing between the scalar triplet and the SM Higgs doublet yields five physical scalar states. Among these, the doubly charged Higgs boson $H^{\pm\pm}$ exhibits distinctive collider signatures due to its unique electric charge and characteristic decay modes. Consequently, $H^{\pm\pm}$ has been a central focus of HTM phenomenology studies \cite{Akeroyd:2005gt,Han:2015hba,Mitra:2016wpr,Du:2018eaw,Li:2018jns}, with particular emphasis on its production at current and future hadron colliders, including the Large Hadron Collider (LHC), the High-Luminosity LHC (HL-LHC) \cite{Apollinari:2015wtw}, and the Future Circular Collider (FCC) \cite{FCC:2018vvp}. The dominant mechanisms for $H^{\pm\pm}$ production are Drell-Yan pair production, associated production with a singly charged Higgs boson, and vector-boson fusion. At the LHC, the pair-production cross section of $H^{\pm\pm}$ ranges from approximately $10^{3}~\mathrm{fb}$ to $10^{-3}~\mathrm{fb}$ as its mass increases from $100~\mathrm{GeV}$ to $1.8~\mathrm{TeV}$ \cite{Fuks:2019clu}. The decay channels of $H^{\pm\pm}$ have also been systematically investigated \cite{FileviezPerez:2008jbu}, with the dominant modes being $H^{\pm\pm} \rightarrow \ell^{\pm} \ell^{\pm}$, $H^{\pm\pm} \rightarrow W^{\pm} W^{\pm}$, and $H^{\pm\pm} \rightarrow H^{\pm}f\bar{f}^{\prime}$ when $m_{H^{\pm\pm}} > m_{H^{\pm}}$ for the normal mass hierarchy.

\par
In addition to the $pp$ collision, other collision modes have also been studied in the literature, including $e^- p$ \cite{Dev:2019hev}, $e^+ e^-$ \cite{Godfrey:2002wp,Agrawal:2018pci}, $e^- e^-$ \cite{Gunion:1995mq}, $e^- \gamma$ \cite{Godfrey:2001xb,Yue:2010zu}, and $\gamma \gamma$ \cite{Chakrabarti:1998qy,Das:2023tna}. These lepton-based collision modes generally operate at lower center-of-mass energies than the $pp$ collision, which limits the accessible parameter space for $m_{H^{\pm\pm}}$. The ATLAS experiment, with assumption of $Br(H^{\pm\pm} \rightarrow \ell^{\pm} \ell^{\pm}) = 1/6$ ($\ell = e,\, \mu,\, \tau$), sets a lower bound of $1080~\mathrm{GeV}$ on $m_{H^{\pm\pm}}$, which exceeds the kinematic reach of current lepton colliders. The Compact Linear Collider (CLIC), proposed to operate at $\sqrt{s} = 1.5~\mathrm{TeV}$ in stage II and $\sqrt{s} = 3.0~\mathrm{TeV}$ in stage III \cite{Aicheler:2018arh}, provides opportunities for probing TeV-scale $H^{\pm\pm}$. Moreover, lepton colliders feature much cleaner backgrounds, therefore, the prospect for discovering $H^{\pm\pm}$ at high-energy lepton colliders could be superior to that at hadron colliders.

\par
In this paper, we systematically investigate the discovery potential of $H^{\pm\pm}$ at CLIC within the HTM in $e^- e^-$, $e^- \gamma$, $\gamma \gamma$ and $e^+ e^-$ collisions, focusing on two representative regions of the parameter space: the Yukawa-like and gauge-like regions. For comparison, we also assess the discovery prospects at the HL-LHC via pair production of the doubly charged Higgs boson. The rest of this paper is organized as follows. In Section \ref{sec:2}, we briefly review the HTM and summarize the theoretical and experimental constraints on the model parameters. Section \ref{sec:3} discusses the dominant production channels of the doubly charged Higgs boson at both CLIC and the LHC. In Section \ref{sec:4}, we analyze the discovery prospects of the doubly charged Higgs at CLIC, while Section \ref{sec:5} presents the corresponding analysis at the HL-LHC for comparison. Finally, a concise summary of our findings is given in Section \ref{sec:6}.

\section{Higgs triplet model}
\label{sec:2}
\par
The Higgs triplet model extends the particle content of the standard model by introducing a complex $SU(2)_L$ triplet scalar field with hypercharge $Y=2$. The scalar triplet $\Delta$ and the SM Higgs doublet $\Phi$ are commonly parameterized as follows:
\begin{equation}
\label{eq:Higgs multiplets}
\Delta
=
\frac{1}{\sqrt{2}}
\begin{pmatrix}
\delta^{+}                                         &   \sqrt{2}\, \delta^{++}
\\
\delta^{0} + i\, \xi^{0} + v_{\Delta}   &   -\, \delta^{+}
\end{pmatrix}\,,
\qquad\qquad
\Phi
=
\frac{1}{\sqrt{2}}
\begin{pmatrix}
\sqrt{2}\, \phi^{+}
\\
\phi^{0} + i\, \eta^{0} + v_{\Phi}
\end{pmatrix}\,,
\end{equation}
where $v_{\Delta}$ and $ v_{\Phi}$ are the vacuum expectation values of the neutral components of $\Delta$ and $\Phi$, respectively.

\par
The Higgs sector of the HTM is governed by the following Lagrangian:
\begin{equation}
\label{LHTM}
\mathcal{L}_{\text{HTM}}
=
\left( D^{\mu}\Phi \right)^{\dag} \left( D_{\mu}\Phi \right)
+
\text{Tr} \Big[ \left( D^{\mu}\Delta \right)^{\dag} \left( D_{\mu}\Delta \right) \Big]
+
\mathcal{L}_{\text{Yukawa}}
- V(\Phi,\, \Delta)\,.
\end{equation}
The covariant derivatives of $\Phi$ and $\Delta$ appearing in the kinetic terms are given by
\begin{equation}
\begin{aligned}
D_{\mu} \Phi
&
= \partial_{\mu} \Phi + i \frac{g_1}{2} V_{1\mu} \Phi + i \frac{g_{2}}{2} V_{2\mu}^{i} \tau^{i} \Phi\,,
\\
D_{\mu} \Delta
&
=
\partial_{\mu} \Delta + i g_{1} V_{1\mu} \Delta + i \frac{g_2}{2} V_{2\mu}^{i} \big[ \tau_{i},\, \Delta \big]\,,
\end{aligned}
\end{equation}
where $V_{1\mu}$ and $V_{2\mu}^{i} ~(i  = 1,\, 2,\, 3)$ denote the $U(1)_{Y}$ and $SU(2)_{L}$ gauge fields, respectively; $g_{1}$ and $ g_{2}$ are the corresponding coupling constants; and $\tau^{i}$ are the Pauli matrices. The Yukawa terms for the Type-II seesaw mechanism are
\begin{equation}
\mathcal{L}_{\text{Yukawa}}
~\supset~
\mathcal{L}_{\text{seesaw}}
=
-\, Y_{ij} \overline{L_{i}^{c}} \,i \tau^{2} \Delta L_{j} + \mathrm{h.c.}\,,
\end{equation}
where $L_i = {\left(\nu_i,\, \ell_i\right)}^T$ are the left-handed lepton doublets, the superscript $c$ denotes the Dirac charge conjugation, and $Y$ is the Yukawa coupling matrix. After electroweak symmetry breaking, the left-handed neutrino states $\nu_{i}$ acquire Majorana masses,
\begin{equation}
m_{\nu} = \sqrt{2} v_{\Delta} Y\,,
\end{equation}
which can be diagonalized by the Pontecorvo–Maki–Nakagawa–Sakata (PMNS) mixing matrix $U$ as
\begin{equation}
m_{\nu} = U^{\ast} \mathrm{diag} \left( m_{1},\, m_{2},\, m_{3} \right) U^{\dagger}\,,
\end{equation}
where $m_{1}$, $m_{2}$ and $m_{3}$ are the masses of the three neutrino mass eigenstates. A detailed discussion on $m_{\nu}$ and $Y$ will be provided in Section \ref{sec:2.2}.

\subsection{Scalar potential and Higgs mass spectrum}
\label{sec:2.1}
\par
The general HTM scalar potential can be written as
\begin{equation}
\begin{aligned}
V(\Phi,\, \Delta)
=
&
- \mu_{\Phi}^{2}\, \Phi^{\dag} \Phi
- \mu_{\Delta}^2\, \text{Tr} \left( \Delta^{\dag} \Delta \right)
+ \frac{\lambda}{4} \left( \Phi^{\dag} \Phi \right)^2
+ \lambda_{1}\, \Phi^{\dag} \Phi\, \text{Tr} \left( \Delta^{\dag} \Delta \right)
\\
&
+ \lambda_{2} \left[ \text{Tr} \left( \Delta^{\dag} \Delta \right) \right]^2
+ \lambda_{3}\, \text{Tr} \left( \Delta^{\dag} \Delta \Delta^{\dag} \Delta \right)
+ \lambda_{4}\, \Phi^{\dag} \Delta \Delta^{\dag} \Phi
\\
&
+ \mu \left( \Phi^T i \tau^{2} \Delta^{\dag} \Phi + \mathrm{h.c.} \right)\,,
\end{aligned}
\end{equation}
where $\mu_{\Phi}$ and $\mu_{\Delta}$ are mass parameters, and $\lambda$ and $\lambda_{1-4}$ are the quartic scalar coupling constants. Using the identity
\begin{equation}
\Phi^{\dag} \Delta \Delta^{\dag} \Phi
+
\Phi^{\dag} \Delta^{\dag} \Delta \Phi
=
\Phi^{\dag} \Phi\, \text{Tr} \left( \Delta^{\dag} \Delta \right)\,,
\end{equation}
the operator $\Phi^{\dag} \Delta^{\dag} \Delta \Phi$ can be expressed in terms of other structures and is thus redundant in the HTM potential. The parameter $\mu$ in the scalar potential is of mass dimension one and explicitly breaks the global $U(1)$ lepton number symmetry. Although $\mu$ is generally complex, it can be rendered real via an appropriate phase rotation of the field combination $\Phi^T i \tau^{2} \Delta^{\dag} \Phi$ \cite{Dey:2008jm,Arhrib:2011uy}.

\par
After EWSB, the Higgs potential minimization conditions imply that the mass parameters can be determined by the Higgs VEVs, the quartic scalar couplings, and the lepton-number-violating trilinear scalar coupling as follows:
\begin{equation}
\begin{aligned}
\mu_{\Phi}^{2}
&=
\frac{1}{4} \lambda v_{\Phi}^{2} + \frac{1}{2} \lambda_{14} v_{\Delta}^{2} - \sqrt{2} \mu v_{\Delta}\,,
\\
\mu_\Delta^{2}
&=
\frac{1}{2} \lambda_{14} v_{\Phi}^{2} + \lambda_{23} v_{\Delta}^{2} - \frac{\mu v_{\Phi}^{2}}{\sqrt{2} v_{\Delta}}\,,
\end{aligned}
\end{equation}
where $\lambda_{ab} \equiv \lambda_{a} + \lambda_{b}~ (a,\, b = 1,\, 2,\, 3,\, 4)$. The Higgs mass eigenstates in the HTM arise from the mixing of the field components given in Eq.\eqref{eq:Higgs multiplets}. The doubly charged scalar fields $\delta^{\pm\pm}$ are already mass eigenstates and are thus also denoted by $H^{\pm\pm}$. The fields $\delta^{\pm}$ and $\phi^{\pm}$ mix to form the singly charged Higgs bosons $H^{\pm}$ and the Goldstone bosons $G^{\pm}$, the latter of which provide the longitudinal degrees of freedom of the $W^\pm$ bosons. Similarly, the mixing of the two $\mathcal{CP}$-odd neutral fields, $\xi^{0}$ and $\eta^{0}$, gives rise to the pseudoscalar $A^0$ and the Goldstone boson $G^0$, with the latter being absorbed by the $Z^0$ boson. Diagonalizing the mass matrix of the $\mathcal{CP}$-even neutral Higgs sector, spanned by $\delta^0$ and $\phi^0$, yields two massive scalar states, $H^0$ and $h^0$. The lighter mass eigenstate, $h^0$, is identified as the $125~ \text{GeV}$ Higgs boson observed in experiments. The rotation angles (also referred to as mixing angles) $\beta_{\pm}$, $\beta_{0}$, and $\alpha$, corresponding respectively to the transformations from the gauge basis to the mass basis for the singly charged, $\mathcal{CP}$-odd neutral, and $\mathcal{CP}$-even neutral Higgs sectors, are given by
\begin{equation}
\tan\beta_{\pm}
=
\frac{1}{\sqrt{2}} \tan\beta_{0}
=
\frac{\sqrt{2} v_{\Delta}}{v_{\Phi}}\,,
\qquad\quad
\tan\alpha
=
-\, \frac{\sqrt{(A - C)^{2} + 4 B^{2}} + (A - C)}{2 B}\,,
\end{equation}
where
\begin{equation}
A = \frac{1}{2} \lambda v_{\Phi}^{2}\,,
\qquad\quad
B = \lambda_{14} v_{\Phi} v_{\Delta} - 2 m_{\Delta}^{2} \frac{v_{\Delta}}{v_{\Phi}}\,,
\qquad\quad
C = m_{\Delta}^2 + 2 \lambda_{23} v_{\Delta}^{2}\,,
\end{equation}
with $m_{\Delta}$ denoting an intermediate mass parameter related to $\mu$ via $m_{\Delta}^{2} = \mu v_{\Phi}^{2}/\sqrt{2} v_{\Delta}$. Accordingly, the scalar mass spectrum of the HTM can be expressed in terms of the parameters $\{ v_{\Phi},\, v_{\Delta},\, m_{\Delta},\, \lambda,\, \lambda_{1-4} \}$ as follows:
\begin{equation}
\begin{aligned}
&
m_{H^{\pm\pm}}^{2}
=
m_{\Delta}^{2} - \lambda_{3} v_{\Delta}^{2} - \frac{\lambda_{4}}{2} v_{\Phi}^{2}\,,
&\qquad&
m_{A^{0}}^{2}
=
m_{\Delta}^{2} \big(1 + \frac{4 v_{\Delta}^{2}}{v_{\Phi}^{2}} \big)\,,
\\
&
m_{H^{\pm}}^{2}
=
\big( m_{\Delta}^{2} - \frac{\lambda_{4}}{4} v_{\Phi}^{2} \big) \big(1 + \frac{2 v_{\Delta}^{2}}{v_{\Phi}^{2}} \big)\,,
&\qquad&
m_{h^{0}, H^{0}}^{2}
=
\frac{1}{2} \left[ (A + C) \mp \sqrt{(A - C)^{2} + 4 B^{2}} \right]\,.
\end{aligned}
\end{equation}

\subsection{Constraints on HTM parameters}
\label{sec:2.2}

\subsubsection{Theoretical constraints on Higgs quartic couplings}
\label{sec:2.2.1}
\par
In the Higgs potential of the HTM, the quartic coupling constants ($\lambda$ and $\lambda_{1,\,2,\,3,\,4}$) are subject to three theoretical constraints: perturbativity, vacuum stability, and perturbative unitarity. Specifically, perturbativity ensures that all couplings remain within the perturbative regime \cite{Haba:2016zbu,Dev:2017ouk}; vacuum stability requires the Higgs potential to be bounded from below in all directions of the field space \cite{Arhrib:2011uy,Chun:2012jw,Bonilla:2015eha,Primulando:2019evb}; and perturbative unitarity constrains scattering amplitudes to remain finite at high energies \cite{Arhrib:2011uy,Dev:2017ouk,Primulando:2019evb}.
\begin{itemize}
\item Perturbativity:
        \begin{equation}
        | \lambda | < 4\pi\,,
        \qquad\quad
        | \lambda_a | < 4\pi
        \quad(a = 1,\, 2,\, 3,\, 4)\,.
        \end{equation}
\item Vacuum stability:
        \begin{equation}
        \begin{aligned}
        &
        \lambda > 0\,,
        \qquad\quad
        \lambda_2 + \lambda_3 > 0\,,
        \qquad\quad
        2 \lambda_2 + \lambda_3 > 0\,,
        \\
        &
        \lambda_1 + \sqrt{\lambda (\lambda_2 + \lambda_3)} > 0\,,
        \qquad\quad
        \lambda_1 + \lambda_4 + \sqrt{\lambda (\lambda_2 + \lambda_3)} > 0\,,
        \\
        &
        \mathrm{max}\big\{\,
        | \lambda_4 | \sqrt{\lambda_2 + \lambda_3} - \lambda_3 \sqrt{\lambda}\,,~
        2 \lambda_1 + \lambda_4 + \sqrt{(2 \lambda - \lambda_4^2/\lambda_3) (2 \lambda_2 + \lambda_3)}
        \,\big\} > 0\,.
        \end{aligned}
        \end{equation}
\item Perturbative unitarity:\footnote{In this work, the parameter $\kappa$ is fixed at $8$.}
        \begin{equation}
        \begin{aligned}
        &
        | \lambda | < 2 \kappa \pi\,,
        \qquad
        | \lambda_1 | < \kappa \pi\,,
        \qquad
        2 | \lambda_2 | < \kappa \pi\,,
        \qquad
        | \lambda_1 + \lambda_4 | < \kappa \pi\,,
        \\
        &
        | 2 \lambda_1 - \lambda_4 | < 2 \kappa \pi\,,
        \qquad
        | 2 \lambda_1 + 3 \lambda_4 | < 2 \kappa \pi\,,
        \qquad
        2 | \lambda_2 + \lambda_3 | < \kappa \pi\,,
        \\
        &
        | 2 \lambda_2 - \lambda_3 | < \kappa \pi\,,
        \\
        &
        \big| \lambda + 4 \lambda_2 + 8 \lambda_3 \pm \sqrt{(\lambda - 4 \lambda_2 - 8 \lambda_3)^2 + 16 \lambda_4^2} \big| < 4 \kappa \pi\,,
        \\
        &
        \big| 3 \lambda + 16 \lambda_2 + 12 \lambda_3 \pm \sqrt{(3 \lambda - 16 \lambda_2 - 12 \lambda_3)^2 + 24 (2\lambda_1 + \lambda_4)^2} \big| < 4 \kappa \pi\,.
        \end{aligned}
        \end{equation}
\end{itemize}

\subsubsection{Experimental constraints on $v_{\Delta}$, Higgs masses and Yukawa couplings}
\label{sec:2.2.2}
\par
In the HTM, the masses of the $W^{\pm}$ and $Z^0$ bosons, arising from the kinetic terms of the Higgs multiplets in Eq.\eqref{LHTM}, are given by
\begin{equation}
m_{W^{\pm}} = \frac{g_2}{2} \sqrt{v_{\Phi}^2 + 2 v_{\Delta}^2}\,,
\qquad\quad
m_{Z^0} = \frac{g_2}{2 \cos\theta_W} \sqrt{v_{\Phi}^2 + 4 v_{\Delta}^2}\,,
\end{equation}
where the Weinberg angle $\theta_W$ is defined as $\cos\theta_W = g_2/\sqrt{g_1^2 + g_2^2}$. The corresponding $\rho$ parameter is then expressed as
\begin{equation}
\rho
\equiv
\frac{m_{W^{\pm}}^2}{m_{Z^0}^2 \cos^2\theta_W}
=
\frac{1 + 2 v_{\Delta}^2/v_{\Phi}^2}{1 + 4 v_{\Delta}^2/v_{\Phi}^2}\,.
\end{equation}
In contrast to the tree-level prediction $\rho = 1$ in the SM, the HTM predicts $\rho < 1$. Electroweak precision measurements yield $\rho = 1.0001 \pm 0.0009$ at the $1\sigma$ confidence level (CL) \cite{ParticleDataGroup:2024cfk}, indicating $v_{\Delta}^2/v_{\Phi}^2 < 4 \times 10^{-4}$. Combined with the relation $\sqrt{v_{\Phi}^2 + 2 v_{\Delta}^2} \simeq 246~ \text{GeV}$, this leads to an upper bound of $v_{\Delta} < 4.9~ \text{GeV}$ for the VEV of the Higgs triplet. In the limit $v_{\Delta} \ll v_{\Phi}$, the scalar spectrum of the HTM can be approximated at leading order by
\begin{equation}
\label{eq:spectrum limit}
\begin{aligned}
&
m_{h^0}^2 = \frac{\lambda}{2} v_{\Phi}^2\,,
&\qquad\quad&
m_{H^{\pm}}^2 = m_{\Delta}^2 - \frac{\lambda_4}{4} v_{\Phi}^2\,,
\\
&
m_{A^0}^2 = m_{H^0}^2 = m_{\Delta}^2\,,
&\qquad\quad&
m_{H^{\pm\pm}}^2 = m_{\Delta}^2 - \frac{\lambda_4}{2} v_{\Phi}^2\,.
\end{aligned}
\end{equation}
The mass of $h^0$ is determined solely by the Higgs doublet's quartic self-coupling $\lambda$ and its VEV $v_{\Phi}$, as in the SM. By contrast, the masses of other exotic Higgs bosons are set by the characteristic mass scale $m_{\Delta}$ of the HTM, with their mass splittings governed by $\lambda_4$. For $\lambda_4 < 0$, the scalar spectrum exhibits a normal mass hierarchy (NMH), $m_{H^{\pm\pm}} > m_{H^{\pm}} > m_{A^0, H^0}$, whereas for $\lambda_4 > 0$, it follows an inverted mass hierarchy (IMH), $m_{H^{\pm\pm}} < m_{H^{\pm}} < m_{A^0, H^0}$.

\par
The existence of a doubly charged Higgs boson is a hallmark prediction of many extensions of the SM, including the Type-II seesaw model, left-right symmetric models, and various radiative neutrino-mass models (e.g., the Zee-Babu model). In proton-proton collisions at the LHC, searches for doubly charged Higgs bosons have primarily focused on Drell-Yan pair and associated production, $q\bar{q} \rightarrow Z^0/\gamma \rightarrow H^{\pm\pm}H^{\mp\mp}$ and $q\bar{q}^{\prime} \rightarrow W^{\pm} \rightarrow H^{\pm\pm}H^{\mp}$, with subsequent fermionic or bosonic decays{\textemdash}manifesting as same-sign dileptons or $W^{\pm}W^{\pm}$ pairs, respectively{\textemdash}determined by the triplet VEV and Yukawa couplings. During Run 2 of the LHC, both ATLAS and CMS analyzed events with same-sign lepton pairs in the two-, three-, and four-lepton final states to search for doubly charged Higgs bosons, which yield relatively clean experimental signatures with low SM backgrounds \cite{ATLAS:2017xqs,ATLAS:2022pbd,ATLAS:2021jol,CMS:2017pet}. No significant excess over the SM predictions was observed; consequently, these searches have set lower limits on the mass of the doubly charged Higgs boson, as summarized in Table \ref{tab:constraint on mH}. It is important to emphasize that the mass limits are highly dependent on several assumptions, including the production mode, chirality, and decay branching fractions. For $H^{\pm\pm}$ decaying predominantly into same-sign light leptons and produced in pairs, the most stringent direct limit from LHC Run 2{\textemdash}{established by the ATLAS measurement using the full $139~ \text{fb}^{-1}$ dataset{\textemdash}is approximately $1.08~ \text{TeV}$. By contrast, for the bosonic decay channel $H^{\pm\pm} \rightarrow W^{\pm}W^{\pm}$, current mass exclusion limits reach only a few hundred GeV{\textemdash}$350~ \text{GeV}$ for pair production and $230~ \text{GeV}$ for associated production{\textemdash}due to more challenging backgrounds.
\begin{table}[htbp]
\centering
\footnotesize
\begin{tabular}{!{\vrule width 1.2pt}c|c|c|ccc!{\vrule width 1.2pt}}
\Hline
\,\textsf{Experiment}\, & \,\textsf{Data sample}\, & \textsf{Benchmark} & \multicolumn{3}{c!{\vrule width 1.2pt}}{\textsf{Observed lower limit [GeV]}}
\\
\Hline
\multirow{9}{*}{ATLAS}
& 
\multirow{5}{*}{\makecell{$36.1~ \text{fb}^{-1}$ \\ \cite{ATLAS:2017xqs}}}
& Pair production & $H_{L}^{\pm\pm}$ & $H_{R}^{\pm\pm}$ &
\\
& & $Br(H^{\pm\pm} \rightarrow e^{\pm}e^{\pm}) = 100\%$ & 768 & 658 &
\\
& & $Br(H^{\pm\pm} \rightarrow e^{\pm}\mu^{\pm}) = 100\%$ & 875 & 761 &
\\
& & $Br(H^{\pm\pm} \rightarrow \mu^{\pm}\mu^{\pm}) = 100\%$ & 846 & 723 &
\\
& & $\sum_{\ell, \ell^{\prime} = e, \mu} Br(H^{\pm\pm} \rightarrow \ell^{\pm}\ell^{\prime \pm}) = 10\%$ & 450 & 320 &
\\
\cline{2-6}
& 
\multirow{4}{*}{\makecell{$139~ \text{fb}^{-1}$ \\ \cite{ATLAS:2022pbd,ATLAS:2021jol}}}
& Pair production & LR model & \multicolumn{2}{l!{\vrule width 1.2pt}}{{\tiny\,}Zee-Babu model}
\\
& & $Br(H^{\pm\pm} \rightarrow \ell^{\pm}\ell^{\prime \pm}) = 1/6,\,\, (\ell, \ell^{\prime} = e, \mu, \tau)$ & 1080 & 900 &
\\
\cline{3-6}
& & Pair and associated production & Pair Prod. & Assoc. Prod. &
\\
& & $Br(H^{\pm\pm} \rightarrow W^{\pm}W^{\pm}) = 100\%$ & 350 & 230 &
\\
\Hline
\multirow{11}{*}{CMS}
& 
\multirow{11}{*}{\makecell{$12.9~ \text{fb}^{-1}$ \\ \cite{CMS:2017pet}}}
& Pair and associated production & Pair Prod. & Assoc. Prod. & Combined
\\
& & $Br(H^{\pm\pm} \rightarrow e^{\pm}e^{\pm}) = 100\%$ & 652 & 734 & 800
\\
& & $Br(H^{\pm\pm} \rightarrow e^{\pm}\mu^{\pm}) = 100\%$ & 665 & 750 & 820
\\
& & $Br(H^{\pm\pm} \rightarrow \mu^{\pm}\mu^{\pm}) = 100\%$ & 712 & 746 & 816
\\
& & $Br(H^{\pm\pm} \rightarrow e^{\pm}\tau^{\pm}) = 100\%$ & 481 & 568 & 714
\\
& & $Br(H^{\pm\pm} \rightarrow \mu^{\pm}\tau^{\pm}) = 100\%$ & 537 & 518 & 643
\\
& & $Br(H^{\pm\pm} \rightarrow \tau^{\pm}\tau^{\pm}) = 100\%$ & 396 & 479 & 535
\\
& & BP1 & 519 & 613 & 723
\\
& & BP2 & 465 & 670 & 716
\\
& & BP3 & 531 & 706 & 761
\\
& & BP4 & 496 & 639 & 722
\\
\Hline
\end{tabular}
\caption{Observed $95\%$ CL lower limits on the $H^{\pm\pm}$ mass from analyses of LHC Run 2 data. In the CMS search using the $12.9~ \text{fb}^{-1}$ dataset, the benchmark points BP1{\textendash}BP4 correspond to four different branching fraction scenarios for $H^{\pm\pm}$ decays, each associated with a distinct neutrino mass hierarchy.}
\label{tab:constraint on mH}
\end{table}

\par
In the HTM, neutrino masses arise from Yukawa interactions between the scalar triplet and the left-handed lepton doublets, with experimental measurements of these masses directly imposing constraints on the Yukawa coupling matrix. Based on $259$ days of data, the KArlsruhe TRItium Neutrino (KATRIN) experiment has set the most recent direct limit on the effective electron antineutrino mass, $m_{\nu} < 0.45~ \text{eV}$, at $90\%$ CL \cite{KATRIN:2024cdt}. In addition, the Planck Collaboration, in combination with baryon acoustic oscillation (BAO) measurements, has established an upper limit on the sum of neutrino masses, $\sum m_{\nu} < 0.12~ \text{eV}$, at $95\%$ CL \cite{Planck:2018vyg}. These complementary constraints provide stringent bounds on neutrino mass scales from both laboratory experiments and cosmological observations. A set of direct and more stringent constraints on the Yukawa coupling matrix is derived from measurements of the branching ratios of lepton-flavor-violating processes \cite{SINDRUM:1987nra,BaBar:2009hkt,HFLAV:2016hnz}, the anomalous magnetic moments of the electron \cite{Hanneke:2008tm} and muon \cite{Muong-2:2006rrc}, the muonium-antimuonium oscillation probability \cite{Willmann:1998gd}, and the Bhabha-scattering cross sections \cite{DELPHI:2005wxt}, all of which have been comprehensively reviewed in the literature \cite{Dev:2017ouk,BhupalDev:2018tox}. In particular, the measurement of Bhabha scattering imposes a direct constraint on the diagonal element $Y_{ee}$, yielding $|Y_{ee}|^2 < 0.12 \times (m_{H^{\pm\pm}}/1\, \text{TeV})^2$. By contrast, constraints from other observables typically involve bilinear combinations of diagonal and off-diagonal Yukawa couplings, implying that sizable diagonal entries must be accompanied by sufficiently small off-diagonal ones. Such a Yukawa texture, however, poses a significant challenge for reproducing the observed PMNS mixing, reflecting the inherent tension between accommodating large diagonal couplings and maintaining the observed neutrino flavor structure. A viable solution is the hybrid seesaw mechanism that incorporates both Type-II and Type-I contributions \cite{Schechter:1980gr}, with neutrino dynamics governed by
\begin{equation}
\mathcal{L}_{\text{seesaw}}^{\text{hybrid}}
=
-\, Y_{ij} \overline{L_{i}^{c}} \,i \tau^{2} \Delta L_{j}
-
g_{ij} \overline{L}_{i} \widetilde{\Phi} N_{j}
-
\frac{1}{2} M_{ij} \overline{N_{i}^{c}} N_{j}
+
\mathrm{h.c.}\,,
\end{equation}
where $N_{i}$ denote the right-handed neutrino singlets. In this hybrid seesaw framework, the Type-I sector shapes the off-diagonal structure of the neutrino mass matrix, while the Type-II contribution primarily determines the diagonal entries, thereby yielding a mass matrix texture that naturally reproduces the observed neutrino mixing pattern and remains compatible with a sizable $Y_{ee}$. In our calculation, we adopt the single-dominance hypothesis, retaining solely the $Y_{ee}$ coupling, which is allowed to approach its experimental upper limit.

\subsection{HTM input parameters}
\label{sec:2.3}
\par
After EWSB, the Higgs potential in the HTM is fully specified by eight parameters, $\{ v_{\Phi},\, v_{\Delta},\, m_{\Delta},\, \lambda,\, \lambda_{1-4} \}$, of which only seven are independent due to the electroweak constraint $\sqrt{v_{\Phi}^2 + 2 v_{\Delta}^2} \simeq 246~ \text{GeV}$. In phenomenological studies, however, it is often more convenient to use relevant physical observables, such as particle masses or mixing angles, as input parameters. In this work, with a focus on the phenomenology of the doubly charged Higgs boson, we adopt $m_{H^{\pm\pm}}$ as one of the fundamental input parameters, while the mass of the lighter $\mathcal{CP}$-even scalar, $m_{h^0}$, is fixed at $125~ \text{GeV}$, corresponding to the experimentally observed Higgs boson. For simplicity, the remaining five independent input parameters of the Higgs potential are chosen as $v_{\Delta}$ and $\lambda_{1-4}$, without reference to the masses of other Higgs states. Notably, such an input scheme effectively avoids the potential fine-tuning problem. As shown in Eq.\eqref{eq:spectrum limit}, the neutral scalars $A^0$ and $H^0$ are degenerate in mass in the small-$v_{\Delta}$ regime, as required by the $\rho$-parameter constraint. Treating both $m_{A^0}$ and $m_{H^0}$ as input parameters inevitably entails fine-tuning of these masses to preserve the perturbativity of the Higgs potential \cite{Du:2018eaw}.

\par
In the Yukawa sector, since the production cross section of the doubly charged Higgs boson at lepton colliders is insensitive to the non-$ee$ elements of the Yukawa coupling matrix, we adopt the single-dominance hypothesis, setting $Y_{ij} = Y_{ee} \delta_{ie} \delta_{je}$. Thus, including the Higgs potential parameters introduced above, the HTM is fully specified by the following seven independent parameters,
\begin{equation}
\{ m_{H^{\pm\pm}},~ v_{\Delta},~ \lambda_{1-4},~ Y_{ee} \}\,,
\end{equation}
with values yet to be determined. A detailed description of the input scheme for these parameters is given below.
\begin{itemize}
\item $\lambda_{1-4}$:
\par
The production and decay channels of the doubly charged Higgs boson studied in this work are independent of the quartic scalar couplings $\lambda_1$ and $\lambda_2$. The dependence on $\lambda_3$ arises exclusively from the $H^{\pm\pm}H^{\mp}H^{\mp}$ Higgs self-coupling. Owing to the strong suppression by the small $v_{\Delta}$, this dependence can be safely neglected; therefore, $\lambda_3$ is set to zero throughout the analysis for simplicity. As shown in Eq.\eqref{eq:spectrum limit}, $\lambda_4$ fully determines the mass hierarchy of the Higgs spectrum. In the NMH scenario considered here, we take $\lambda_4 = -2$ as a representative value consistent with both theoretical and experimental constraints, reflecting a typical mass-splitting pattern.
\item $Y_{ee}$, $v_{\Delta}$ and $m_{H^{\pm\pm}}$:
\par
The decay patterns of the doubly charged Higgs boson are primarily determined by its mass, the mass splitting between the doubly and singly charged Higgs states, and the relative strength of Yukawa and gauge couplings of the Higgs triplet. Figure \ref{fig1} illustrates the branching ratios of the dominant decay channels of the doubly charged Higgs boson as functions of $v_{\Delta}$ for various Higgs masses, assuming $\sqrt{2} v_{\Delta} Y_{ee} = 0.1~ \text{eV}$. As $m_{H^{\pm\pm}}$ increases from $400~ \text{GeV}$ to $2500~ \text{GeV}$, the mass splitting, $\Delta m = m_{H^{\pm\pm}} - m_{H^{\pm}}$, gradually narrows from approximately $40~ \text{GeV}$ to about $6~ \text{GeV}$. In the small-$v_{\Delta}$ regime, the decays of the doubly charged Higgs are dominated by Yukawa interactions, proceeding primarily into same-sign dilepton final states. As $v_{\Delta}$ increases, the three-body decays into a singly charged Higgs and a pair of light fermions become increasingly significant. At sufficiently large $v_{\Delta}$, the decays are governed by gauge interactions, rendering the same-sign $W$-pair channel the leading mode. Although the NMH scenario permits decays into singly charged Higgs bosons, we confine our analysis to the Yukawa-like and gauge-like regions, explicitly excluding the intermediate-$v_{\Delta}$ regime; in these two limits, $H^{\pm\pm}$ decays almost exclusively into $e^{\pm}e^{\pm}$ and $W^{\pm}W^{\pm}$, with $Br(H^{\pm\pm} \rightarrow e^{\pm}e^{\pm}) \simeq 100\%$ and $Br(H^{\pm\pm} \rightarrow W^{\pm}W^{\pm}) \simeq 100\%$, respectively. Unless otherwise specified, the subsequent analysis is conducted at the following two extreme benchmark points in the $(v_{\Delta}, Y_{ee})$ parameter space:
\begin{equation}
\begin{aligned}
&
\text{BP1}:
&\quad&
(\,v_{\Delta} = 2 \times 10^{-10}~ \text{GeV}\,,
\quad~~
Y_{ee} = 0.35\,)
&&
\in
&~& \text{Yukawa-like region},
\\
&
\text{BP2}:
&\quad&
(\,v_{\Delta} = 4.5~ \text{GeV}\,,
\quad~
Y_{ee} = 1.5 \times 10^{-11}\,)
&&
\in
&~&
 \text{gauge-like region},
\end{aligned}
\end{equation}
which correspond, respectively, to values of $Y_{ee}$ and $v_{\Delta}$ close to their experimental upper limits. In light of the constraints from LHC measurements, the scan range of $m_{H^{\pm\pm}}$ is set to be
\begin{equation}
m_{H^{\pm\pm}} \,\geqslant\,
\left\lbrace~
\begin{aligned}
&
1100~ \text{GeV}
&\quad~&
\text{(Yukawa-like region)}\,,
\\
&
\,400~\; \text{GeV}
&\quad~&
\text{(gauge-like region)}\,.
\end{aligned}
\right.
\end{equation}
\end{itemize}
\begin{figure}[htbp]
\centering
\begin{minipage}{0.9\textwidth}
\includegraphics[width=1.0\textwidth]{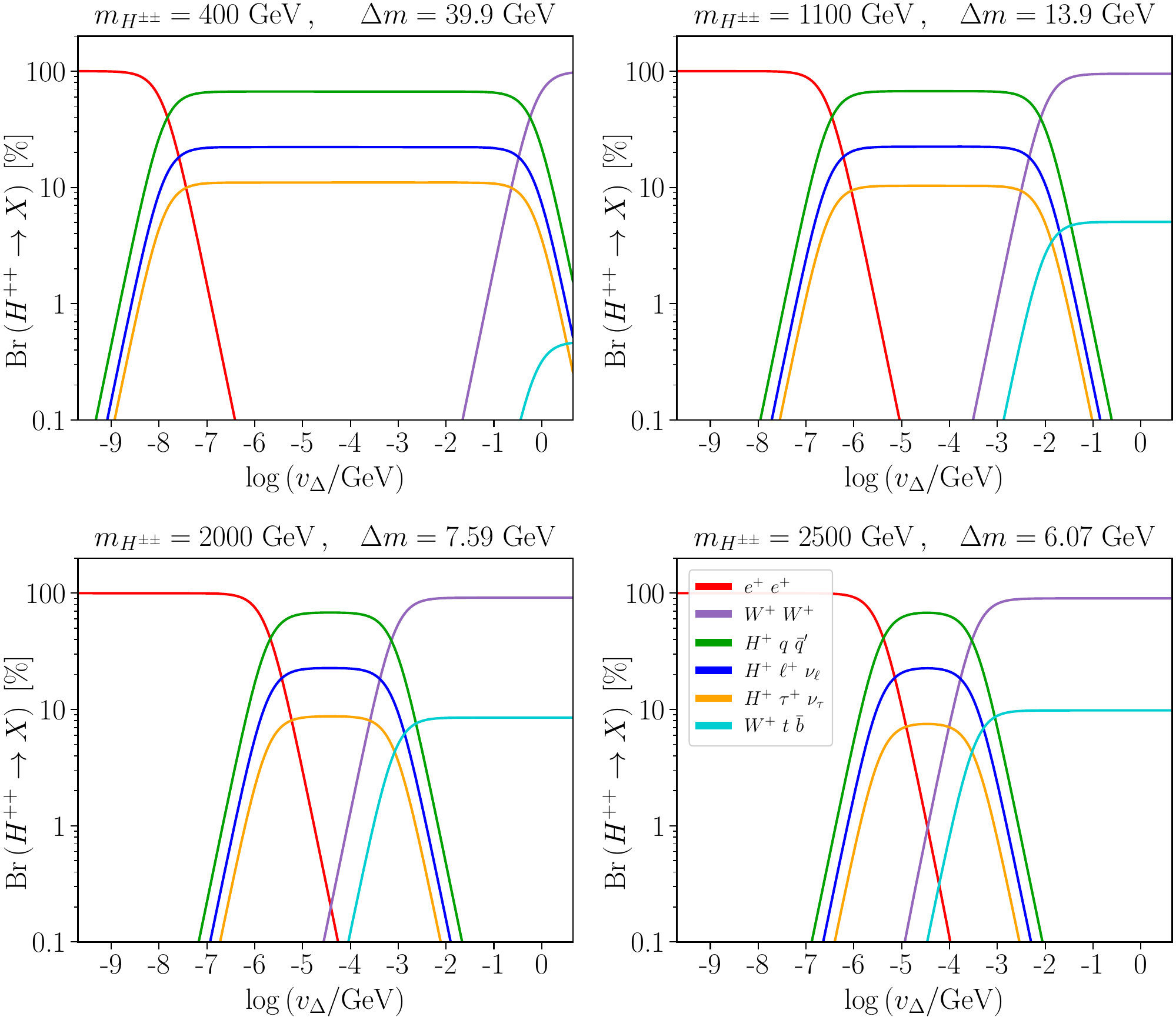}
\end{minipage}
\caption{Branching ratios of the dominant decay channels of $H^{++}$ versus $v_{\Delta}$ for different Higgs masses, with $\sqrt{2} v_{\Delta} Y_{ee} = 0.1~ \text{eV}$.}
\label{fig1}
\end{figure}

\section{Doubly charged Higgs production}
\label{sec:3}
\par
Based on the HTM input parameters discussed above, we identify the dominant production mechanisms of $H^{\pm\pm}$ at lepton colliders in four different collision modes: $e^+e^-$, $e^-e^-$, $e^-\gamma$, and $\gamma\gamma$. Among these, the $e^+e^-$ mode is typically the primary operating mode at lepton colliders, offering the highest integrated luminosity. The high-energy photon beams in the initial state are generated via Compton backscattering, with the energy spectrum given by \cite{Ginzburg:1981vm}
\begin{equation}
\label{eq:photonpdf}
\frac{1}{\sigma_c} \frac{d \sigma_c}{d y}
=
\frac{2\sigma_0}{x\sigma_c} \left[ 1-y+ \frac{1}{1-y} - \frac{4y}{x(1-y)} + \frac{4 y^2}{x^2 {(1-y)}^2} \right] ,
\end{equation}
where $\sigma_c$ denotes the total Compton scattering cross section, and $y$ represents the fraction of the scattered photon energy relative to the incident electron-beam energy. The kinematic upper bound on $y$ is $y_{\text{max}} = x/(1+x)$, with $x$ being a dimensionless parameter determined by the laser-electron configuration. In this work, we adopt $x = 4.8$ \cite{Boudjema:1998sw,Ginzburg:2019yws}, which corresponds to $y_{\text{max}} \simeq 0.8$. As a result, the maximal center-of-mass energies of the $\gamma\gamma$ and $e^-\gamma$ collision modes are approximately 80\% and 90\% of that of the $e^+ e^-$ mode, respectively. For event generation, the HTM is implemented using \texttt{FeynRules} \cite{Alloul:2013bka} to derive the relevant Feynman rules, which are subsequently interfaced with \texttt{MadGraph5\_aMC@NLO} \cite{Alwall:2014hca} to generate $H^{\pm\pm}$ events.

\begin{itemize}
\item \textbf{$e^+ e^-$ mode}: \\
For $\sqrt{s} > 2 m_{H^{\pm\pm}}$, the production of $H^{\pm\pm}$ is dominated by the Drell-Yan pair production, which is independent of $v_{\Delta}$ but exhibits mild sensitivity to $Y_{ee}$. When $\sqrt{s}$ falls below the $H^{\pm\pm}$ pair-production threshold, the dominant production mechanisms shift to $e^+ e^- \to H^{\pm\pm} e^{\mp} e^{\mp}$ and $e^+ e^- \rightarrow H^{\pm\pm} W^{\mp} W^{\mp}$. Both processes depend on $Y_{ee}$, but only the latter is sensitive to $v_{\Delta}$.

\item \textbf{$e^- e^-$ mode}: \\
The leading channels in the $e^- e^-$ collision mode are $e^- e^- \rightarrow H^{--} \gamma$ and $e^- e^- \rightarrow H^{--} \nu_e \nu_e$. The former channel shows no dependence on $v_{\Delta}$ but scales strongly with $Y_{ee}$, rendering it the dominant production channel in the Yukawa-like region. By contrast, the latter channel is sensitive to both $v_{\Delta}$ and $Y_{ee}$, dominating in the gauge-like region. For the benchmark parameters adopted in this analysis, the $H^{--} \nu_e \nu_e$ cross section is significantly smaller than that of the photon-associated channel.

\item \textbf{$e^- \gamma$ mode}: \\
In the Yukawa-like region, the process $e^- \gamma \rightarrow H^{--} e^+$ dominates the production of $H^{--}$, with a cross section unaffected by $v_{\Delta}$ and governed primarily by $Y_{ee}$. In the gauge-like region, the two principal production channels are $e^- \gamma \rightarrow H^{++} H^{--} e^-$ and $e^- \gamma \rightarrow H^{--} W^+ \nu_e$. For the benchmark scenario under consideration, the cross sections of these two channels are significantly smaller than that of $H^{--} e^+$ production in the Yukawa-like region.

\item \textbf{$\gamma \gamma$ mode}: \\
In this collision mode, $H^{\pm\pm}$ is predominantly produced via the Drell-Yan process when kinematically allowed, with a cross section independent of both $v_{\Delta}$ and $Y_{ee}$. Below the $H^{\pm\pm}$ pair-production threshold, the doubly charged Higgs is primarily produced via $\gamma\gamma \rightarrow H^{\pm\pm} e^{\mp} e^{\mp}$ in the Yukawa-like region, with a cross section that depends on $Y_{ee}$ but is independent of $v_{\Delta}$. By contrast, in the gauge-like region, production mainly proceeds via $\gamma\gamma \rightarrow H^{\pm\pm} W^{\mp} W^{\mp}$, which is governed by $v_{\Delta}$ and insensitive to $Y_{ee}$.
\end{itemize}

\par
For comparison, we also consider the dominant production mechanisms of the doubly charged Higgs boson at the LHC, including
\begin{equation}
\begin{aligned}
  & p p \rightarrow H^{\pm\pm} H^{\mp} + X\,, & \qquad \quad & p p \rightarrow H^{++} H^{--} + X\,, \\
  & p p \rightarrow H^{\pm\pm} j j + X\,,     & \qquad \quad & p p \rightarrow H^{\pm\pm} W^{\mp} + X\,.
\end{aligned}
\end{equation}
Among these, the associated production with a singly charged Higgs and the Drell-Yan pair production are independent of $v_{\Delta}$ and $Y_{ee}$, and constitute the leading production modes at the LHC. The other two processes exhibit a pronounced sensitivity to $v_{\Delta}$ and can attain sizable cross sections in the gauge-like region. In summary, all relevant production channels along with their dependence on $v_{\Delta}$ and $Y_{ee}$ are listed in Table \ref{tab:vD-Yee-dep}.

\begin{table}[htbp]
\centering
\setlength{\tabcolsep}{8pt}
\begin{tabular}{!{\vrule width 1.2pt}c|c|c|c!{\vrule width 1.2pt}c|c|c|c!{\vrule width 1.2pt}}
\Hline
\textsf{Initial State} & \textsf{Final State} & \depbox{\textsf{$v_{\Delta}$}} & \depbox{\textsf{$Y_{ee}$}} &
\textsf{Initial State} & \textsf{Final State} & \depbox{\textsf{$v_{\Delta}$}} & \depbox{\textsf{$Y_{ee}$}} \\
\Hline
\multirow{3}{*}{$e^{+} e^{-}$}
& $H^{++}\ H^{--}$ & & \checkmark &
\multirow{3}{*}{$e^- e^-$}
& $H^{--}\ \gamma$ & & \checkmark \\
& $H^{\pm\pm}\ e^\mp\ e^\mp$ & & \checkmark &
& $H^{--}\ \nu_e\ \nu_e$ & \checkmark & \checkmark \\
& $H^{\pm\pm}\ W^\mp\ W^\mp$ & \checkmark & \checkmark & & & & \\
\Hline
\multirow{3}{*}{$e^- \gamma$}
& $H^{--}\ e^+$ & & \checkmark &
\multirow{3}{*}{$\gamma\hspace{0.49em} \gamma$}
& $H^{++}\ H^{--}$ & & \\
& $H^{++}\ H^{--}\ e^-$ & & \checkmark &
& $H^{\pm\pm}\ e^\mp\ e^\mp$ & & \checkmark \\
& $H^{--}\ W^+\ \nu_e$ & \checkmark & \checkmark &
& $H^{\pm\pm}\ W^\mp\ W^\mp$ & \checkmark & \\
\Hline
\multirow{2}{*}{$p\ p$}
& $H^{\pm\pm}\ H^\mp$ & & &
\multirow{2}{*}{$p\ p$}
& $H^{\pm\pm}\ j\ j$ & \checkmark & \\
& $H^{++}\ H^{--}$ & & &
& $H^{\pm\pm}\ W^\mp$ & \checkmark & \\
\Hline
\end{tabular}
\caption{\label{tab:vD-Yee-dep}
Dominant production channels of $H^{\pm\pm}$ and their dependence on $v_{\Delta}$ and $Y_{ee}$ at lepton and hadron colliders.}
\end{table}

\par
Figure \ref{fig2} presents the production cross sections of the doubly charged Higgs boson as functions of its mass at the $3~\mathrm{TeV}$ CLIC and the $14~\mathrm{TeV}$ LHC. The two panels on the left illustrate the dominant production channels at BP1, where the production cross sections at CLIC are generally much larger than at the LHC. At CLIC, the processes $e^- e^- \rightarrow H^{--} \gamma$ and $e^- \gamma \rightarrow H^{--} e^+$ yield the largest production cross sections. The former's cross section increases with $m_{H^{\pm\pm}}$ due to enhancement from the $s$-channel propagator $1/(s-m_{H^{\pm\pm}}^2)$, whereas the latter's decreases with increasing $m_{H^{\pm\pm}}$ primarily due to phase-space suppression. Pair-production channels also yield appreciable cross sections, but they diminish rapidly with increasing $m_{H^{\pm\pm}}$ owing to kinematic threshold effects. Overall, across most of the mass range, the production cross sections at CLIC are at least one order of magnitude higher than those at the LHC. The two panels on the right of Fig.\ref{fig2} display the dominant production channels at BP2, where $H^{\pm\pm}$ pair production becomes the leading mechanism at CLIC. Nevertheless, its cross section is strongly constrained by kinematic thresholds and therefore decreases rapidly with increasing $m_{H^{\pm\pm}}$.

Based on the hierarchical structure of the production cross sections, we identify representative channels at CLIC for each benchmark point, BP1 and BP2,
\begin{equation}
\begin{aligned}
&
\text{BP1}:
&~~&
e^- e^- \rightarrow H^{--} \gamma\,,
&\qquad&
e^- \gamma \rightarrow H^{--} e^+\,,
\\
&
\text{BP2}:
&~~&
\gamma \gamma \rightarrow H^{++} H^{--}\,,
&\qquad&
e^+ e^- \rightarrow H^{++} H^{--}\,.
 \end{aligned}
\end{equation}
which will be studied in detail in Sec.\ref{sec:4}. Moreover, we will analyze $H^{\pm\pm}$ pair production at the $14~\mathrm{TeV}$ LHC for both benchmark scenarios in Sec.\ref{sec:5} to compare the discovery potential of the doubly charged Higgs boson at different colliders.

\begin{figure}[htbp]
\centering
\begin{minipage}{0.9\textwidth}
\includegraphics[width=1.0\textwidth]{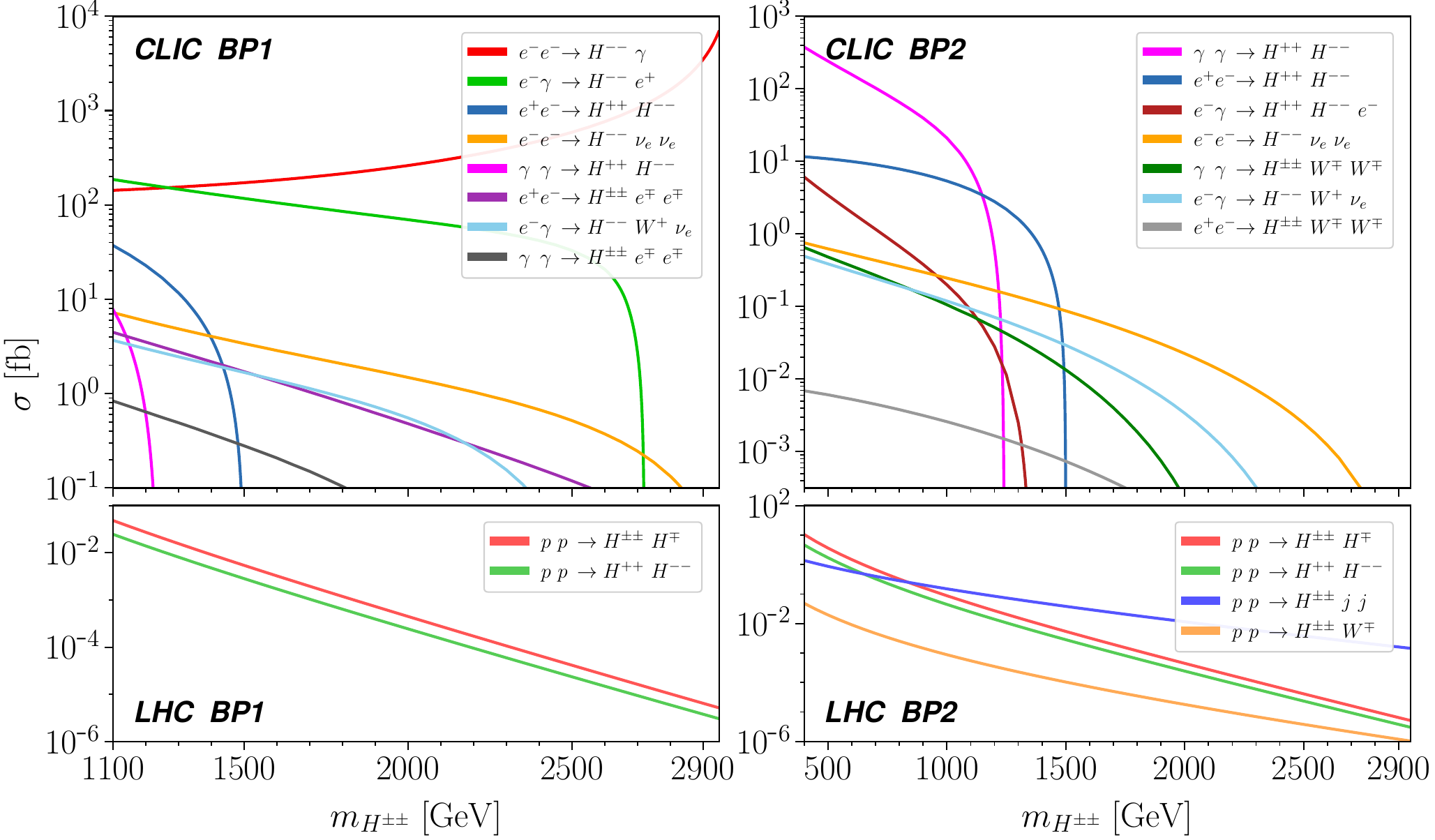}
\end{minipage}
\caption{Production cross sections of $H^{\pm\pm}$ as functions of $m_{H^{\pm\pm}}$ at the $3~\mathrm{TeV}$ CLIC and the $14~\mathrm{TeV}$ LHC for both BP1 and BP2.}
\label{fig2}
\end{figure}

\section{Discovery potential at CLIC}
\label{sec:4}
In this section, we present a detailed study of the discovery potential for doubly charged Higgs bosons at CLIC. Since the projected center-of-mass energy at CLIC Stage I is only 380~GeV, our analysis primarily exploits the data from CLIC Stage II ($\sqrt{s} = 1.5~\mathrm{TeV}$) and Stage III ($\sqrt{s} = 3.0~\mathrm{TeV}$), with integrated luminosities of $2.5~\mathrm{ab}^{-1}$ and $5.0~\mathrm{ab}^{-1}$, respectively \cite{Brunner:2022usy}. These values refer to the $e^+ e^-$ collision mode; for the other three collision modes, we assume integrated luminosities of $25~\mathrm{fb}^{-1}$ at Stage II and $50~\mathrm{fb}^{-1}$ at Stage III.

\par
Both signal and background events are generated using \texttt{MadGraph5\_aMC@NLO} \cite{Alwall:2014hca}, with the decays of unstable particles handled by \texttt{MadSpin} \cite{Artoisenet:2012st}. Initial- and final-state radiation effects are simulated with \texttt{Pythia8} \cite{Bierlich:2022pfr}. The detector effects, such as tracking efficiency and energy/momentum resolution, are incorporated through \texttt{Delphes} \cite{deFavereau:2013fsa} employing the CLIC detector configuration \cite{Leogrande:2019qbe}. For lepton (electrons or muons, unless otherwise stated) and photon reconstruction, an isolation criterion is applied: the scalar sum of $p_T$ for particles within $\Delta R = 0.1$ around the reconstructed object must be less than 20\% of its $p_T$. Final-state jets are reconstructed via the \texttt{FastJet} \cite{Cacciari:2011ma} package, clustered with the Valencia Linear Collider (VLC) algorithm \cite{Boronat:2014hva,Boronat:2016tgd} using inclusive mode. If a $b$-jet is present in the final state, the $70\%$ efficiency working point is employed \cite{Leogrande:2019qbe}. The beam-induced backgrounds (e.g., $\gamma\gamma \rightarrow \text{hadrons}$) are simulated by applying additional energy smearing to the reconstructed jets. For the background analysis, the following baseline selection criteria are defined for the final-state objects:
\begin{equation}
\begin{aligned}
\label{eq:bs}
&
p_{T, \ell} > 10~\mathrm{GeV}\,,
&\quad~&
p_{T, \gamma} > 10~\mathrm{GeV}\,, 
&\quad~&
p_{T, j} > 20~\mathrm{GeV}\,,
\\
&
\abs{\eta_{\ell}} < 2.5\,, 
&\quad~&
\abs{\eta_{\gamma}} < 2.5\,,
&\quad~&
\abs{\eta_{j}} < 5.0\,.
\end{aligned}
\end{equation}
The event selection is implemented in \texttt{MadAnalysis5} \cite{Conte:2012fm}. As illustrated in Fig.\ref{fig1}, $H^{\pm\pm}$ decays predominantly into same-sign lepton pairs in the Yukawa-like region, and into same-sign $W$-boson pairs in the gauge-like region. In the background analysis, we adopt the following convention: if a scattering process involves an intermediate state, only the resonance associated with that state is considered; otherwise, it represents the residual non-resonant contribution after all resonances have been subtracted.

\subsection{$e^- e^- \rightarrow H^{--} \gamma \rightarrow e^- e^- \gamma$ at BP1}
\label{sec:4.1}
\par
As shown in the top-left panel of Fig.\ref{fig2}, $H^{--}\gamma$ associated production via $e^-e^-$ collision is the dominant production channel in the Yukawa-like region. We therefore prioritize the study of this production mechanism along with the subsequent decay $H^{--} \rightarrow e^- e^-$. The signal is required to contain two electrons and one photon in the final state, all of which must pass the baseline selection criteria. The major SM backgrounds include $e^- e^- \rightarrow e^- e^- \gamma$ and $e^- e^- \rightarrow W^- e^- \nu_e \gamma \rightarrow e^- \bar{\nu}_e e^- \nu_e \gamma$, with the former contributing approximately $95\%$ of the total background. Table \ref{tab:ee-sgnf} summarizes the numbers of signal and background events after the baseline selection, together with the corresponding statistical significances $\mathcal{S}$, defined as
\begin{equation}
\mathcal{S} = \frac{N_S}{\sqrt{N_S + N_B}}\,,
\end{equation}
where $N_S$ and $N_B$ denote the expected numbers of signal and background events, respectively. For each representative value of $m_{H^{\pm\pm}}$, the expected numbers of signal and background events are both approximately of $\mathcal{O}(10^3)$, resulting in a statistical significance well above $5\sigma$ and thus obviating the need for further kinematic cuts. As the signal cross section scales with $Y_{ee}^2$, the minimum $Y_{ee}$ required to achieve a $5\sigma$ significance can be inferred. For each $m_{H^{\pm\pm}}$, this value lies well below the current upper limit from  Bhabha scattering experiments, $Y_{ee} < 0.35$. For the Yukawa coupling in the range $Y_{ee} \in [0.05,\, 0.11]$, the doubly charged Higgs boson with mass between 1100 and $2500~\mathrm{GeV}$ could be discovered at the future CLIC. This demonstrates that the $e^-e^-$ mode at CLIC provides excellent sensitivity for probing the doubly charged Higgs boson when it predominantly decays into a same-sign lepton pair.

\begin{table}[htbp]
  \centering
  \setlength{\tabcolsep}{7pt}
  \begin{tabular}{!{\vrule width 1.2pt} c | c | c c c c!{\vrule width 1.2pt}}
    \Hline
    $\sqrt{s}~[\mathrm{TeV}]$ & $1.5$ & \multicolumn{4}{c!{\vrule width 1.2pt}}{\hspace{-0.7em}$3.0$} \\
    $m_{H^{\pm\pm}}~[\mathrm{GeV}]$ & $1100$ & $1100$ & $1500$ & $2000$ & $2500$  \\
    \Hline
    $N_{\text{signal}} \,/\, 10^3$ & $22.4$ & $4.38$ & $5.42$ & $8.40$ & $19.0$ \\
    $N_{\text{bkg}} \,/\, 10^3$ & $11.8$ & \multicolumn{4}{c!{\vrule width 1.2pt}}{\hspace{-0.8em}$7.49$} \\
    $\mathcal{S}$ & $121$ & $40.2$ & $47.7$ & $66.7$ & $117$ \\
    \Hline
    $Y_{ee}$ & $0.0551$ & $0.112$ & $0.100$ & $0.0806$ & $0.0536$ \\
    \Hline
  \end{tabular}
  \caption{
  \label{tab:ee-sgnf}
  Event numbers for the signal process $e^- e^- \rightarrow H^{--} \gamma \rightarrow e^- e^- \gamma$ and its background, along with the corresponding statistical significance and the minimum $Y_{ee}$ required for a $5\sigma$ discovery.}
\end{table}

\subsection{$e^- \gamma \rightarrow H^{--} e^+ \rightarrow e^- e^- e^+$ at BP1}
\label{sec:4.2}
In the Yukawa-like region, the process $e^- \gamma \rightarrow H^{--} e^+$ represents another major production channel for the doubly charged Higgs boson, with a cross section comparable to that of $e^- e^- \to H^{--} \gamma$ at low masses. For the signal process $e^- \gamma \rightarrow H^{--} e^+ \rightarrow e^- e^- e^+$, the event selection requires exactly three final-state electrons with a total electric charge of $-e$, all of which must satisfy the baseline selection criteria. The dominant SM backgrounds are
\begin{equation}
\label{eq:background-B}
\begin{aligned}
&
(1)
&&
e^+ e^- e^-\,,
&\quad~&
(2)
&&
W^+ W^- e^-\,,
&\quad~&
(3)
&&
W^- Z \nu_e\,,
&\quad~&
(4)
&&
Z e^-\,.
\end{aligned}
\end{equation}
Among them, the irreducible process $e^- \gamma \rightarrow e^+ e^- e^-$ is the dominant contribution, accounting for approximately $70\%$ of the total background. The processes $e^-\gamma \rightarrow W^+ W^- e^-$ and $W^- Z \nu_e$ lead to the same final state as the signal via leptonic decays of the weak gauge bosons and constitute subleading backgrounds. By contrast, the $Z e^-$ background contributes the least, accounting for about $5\%$ of the total. Notably, the process $e^- \gamma \rightarrow e^- j j$ has a sizable production cross section, and may therefore be regarded as a SM background if the final-state jets are misidentified as an oppositely charged electron pair. However, with the electron isolation criterion applied, the probability of a jet being misidentified as an electron is well below $\mathcal{O}(10^{-3})$, rendering this background negligible compared to those listed in Eq.\eqref{eq:background-B}. 

\par
After applying only the baseline selection criteria, the signal significance is sufficiently high, and no additional kinematic cuts are necessitated. Table \ref{tab:ea-sgnf} presents the event yields for the signal and background, along with the corresponding signal significances, at CLIC Stage II and Stage III after the baseline selection. For each selected value of $m_{H^{\pm \pm}}$, both the signal and background yield roughly $10^3$ events, resulting in a statistical significance comfortably exceeding the $5\sigma$ threshold. The table also lists the minimal $Y_{ee}$ required to reach a $5\sigma$ observation within the mass range $[1100,\, 2500]~\mathrm{GeV}$, which in all cases lies below the current experimental upper limit. These results suggest that the channel $e^- \gamma \rightarrow H^{--} e^+ \rightarrow e^- e^- e^+$ constitutes a promising and sensitive pathway for discovering the doubly charged Higgs boson when the Yukawa coupling to charged leptons is sufficiently large.

\begin{table}[htbp]
  \centering
  \setlength{\tabcolsep}{7pt}
  \begin{tabular}{!{\vrule width 1.2pt} c | c | c c c c!{\vrule width 1.2pt}}
    \Hline
    $\sqrt{s}~[\mathrm{TeV}]$ & $1.5$ & \multicolumn{4}{c!{\vrule width 1.2pt}}{\hspace{0.5em} $3.0$} \\
    $m_{H^{\pm\pm}}~[\mathrm{GeV}]$ & $1100$ & $1100$ & $1500$ & $2000$ & $2500$ \\
    \Hline
    $N_{\text{signal}} \,/\, 10^3$ & $3.42$ & $5.56$ & $3.65$ & $2.21$ & $1.03$ \\
    $N_{\text{bkg}} \,/\, 10^3$ & $1.82$ & \multicolumn{4}{c!{\vrule width 1.2pt}}{\hspace{0.2em} $1.23$} \\
    $\mathcal{S}$ & $47.2$ & $67.5$ & $52.2$ & $37.7$ & $21.7$ \\
    \Hline
    $Y_{ee}$ & $0.0900$ & $0.0644$ & $0.0795$ & $0.102$ & $0.150$ \\
    \Hline
  \end{tabular}
  \caption{\label{tab:ea-sgnf}
  Same as Table \ref{tab:ee-sgnf}, but for $e^- \gamma \rightarrow H^{--} e^+ \rightarrow e^- e^- e^+$.}
\end{table}

\subsection{$\gamma \gamma \rightarrow H^{++} H^{--} \rightarrow W^+ W^+ W^- W^-$ at BP2}
\label{sec:4.3}
In the gauge-like region, the most dominant production mechanism for the doubly charged Higgs boson is pair production via $\gamma\gamma$ collisions, whenever kinematically allowed, as illustrated in the top-right panel of Fig.\ref{fig2}. The produced $H^{\pm\pm}$ bosons subsequently decay exclusively into same-sign $W$-boson pairs, leading to the signal process $\gamma \gamma \rightarrow H^{++} H^{--} \rightarrow W^+ W^+ W^- W^-$. Taking into account the $W$-boson decay branching fractions and the complexity of the relevant backgrounds, we focus on signal events featuring a same-sign dilepton (SSDL) in the final state, specifically $4W \rightarrow \ell^{\pm}\ell^{\pm} + 2\nu + 4q$. To optimize the signal selection based on the final-state topology, we require events to contain two same-sign leptons and at least three jets{\textemdash}excluding $b$-jets to suppress top-related backgrounds{\textemdash}i.e., $\ell^\pm \ell^\pm + \geq 3j$. This semi-leptonic topology offers a favorable balance: it yields significantly higher statistics than the fully leptonic channel while avoiding the overwhelming multijet backgrounds associated with the fully hadronic channel. It should be noted that the decay $4W \rightarrow \ell^{+}\ell^{-}\ell^{\pm} + 3\nu + 2q$ also contributes to the signal final state, accounting for approximately $10\%$ of the total signal events. We identify the following dominant backgrounds for the $\ell^\pm \ell^\pm + \geq 3j$ final state:
\begin{equation}
\begin{aligned}
&
(1)
&&
4W\,,
&\quad&
(2)
&&
W W Z\,,
&\quad&
(3)
&&
t\bar{t}\,,
&\quad&
(4)
&&
 t b W\,,
 &\quad&
 (5)
 &&
 W W h\,.
\end{aligned}
\end{equation}
The first two correspond to prompt lepton backgrounds, in which SSDLs arise directly from vector boson decays. Contributions from both the $2\ell$ and $3\ell$ final states are included to ensure consistency with the signal treatment. These two processes form the dominant background. For $t\bar{t}$ and $tbW$ backgrounds, the SSDL primarily consists of a prompt lepton from a $W$-boson decay and a non-prompt lepton originating from a $b$-hadron decay. For the $WWh$ background, contributions originate from either prompt production via $h \rightarrow VV^{\ast}$ decays or non-prompt production via the $h \rightarrow b\bar{b}$ decay. The related information on the signal and background events is provided in Table \ref{tab:aa-sgnf}. In the low-mass regime ($m_{H^{\pm\pm}} \lesssim 600~\mathrm{GeV}$), the statistical significance substantially surpasses the $5\sigma$ discovery threshold. Given that the signal significance is relatively low in the high-mass region, a set of kinematic cuts is introduced to better discriminate the signal from the background.

\begin{table}[htbp]
  \centering
  \setlength{\tabcolsep}{7pt}
  \begin{tabular}{!{\vrule width 1.2pt} c | c | c c c c!{\vrule width 1.2pt}}
  \Hline
    $\sqrt{s}~[\mathrm{TeV}]$ & $1.5$ & \multicolumn{4}{c!{\vrule width 1.2pt}}{$3.0$} \\
    $m_{H^{\pm\pm}}~[\mathrm{GeV}]$ & $400$ & $400$ & $600$ & $800$ & $1000$ \\
  \Hline
    $N_{\text{signal}}$ & $275$ & $664$ & $279$ & $95.0$ & $23.9$ \\
    $N_{\text{bkg}}$ & $49.1$ & \multicolumn{4}{c!{\vrule width 1.2pt}}{$202$} \\
    $\mathcal{S}$ & $15.3$ & $22.6$ & $12.7$ & $5.51$ & $1.59$ \\
  \Hline
  \end{tabular}
  \caption{\label{tab:aa-sgnf}
  Event numbers for the signal $\gamma \gamma \rightarrow H^{++}H^{--} \rightarrow 4W \rightarrow \ell^{\pm} \ell^{\pm} + \geq 3j$ and its background, along with the corresponding statistical significance.}
\end{table}

\par
Figure \ref{fig3} shows selected kinematic distributions of the $\ell^{\pm}\ell^{\pm} + \geq 3j$ final state at $\sqrt{s} = 3~ \mathrm{TeV}$ CLIC after the baseline selection, including the transverse momentum and the pseudorapidity of the leading lepton, the invariant mass of the three leading jets, and the $H_T$ variable, defined as
\begin{equation}
H_T = \sum_{j \in \text{jets}} p_{T,\, j} +\, \slashed{p}_{T}\,,
\end{equation}
where the scalar sum runs over all reconstructed jets and $\slashed{p}_T$ denotes the missing transverse momentum. As illustrated in the figure, the signal and background can be well separated in these kinematic distributions. To improve the signal significance in the high-mass region, we therefore apply the following kinematic cuts:
\begin{equation}
\label{eq:optimized-selection}
p_{T,\, \ell_1} > 120~\mathrm{GeV}\,, \quad~ \abs{\eta_{\ell_1}} < 1.5\,, \quad~ M_{jjj} > 500~\mathrm{GeV}\,, \quad~ H_T > 800~\mathrm{GeV}\,.
\end{equation}
\begin{figure}[htbp]
\centering
\begin{minipage}{0.45\textwidth}
\includegraphics[width=1.0\textwidth]{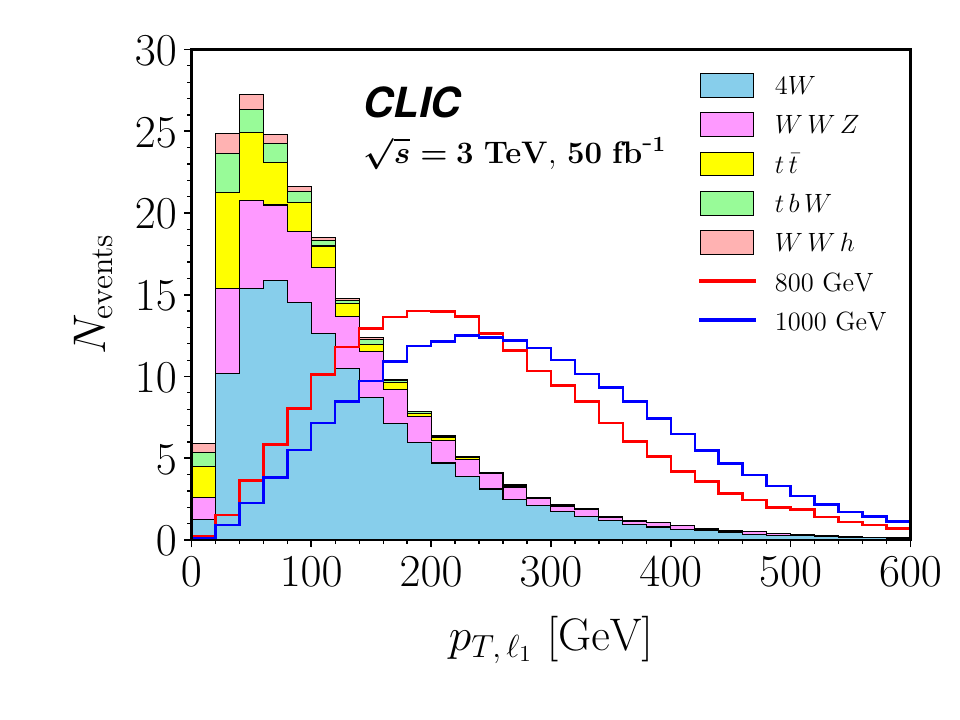}
\end{minipage}
\begin{minipage}{0.45\textwidth}
\includegraphics[width=1.0\textwidth]{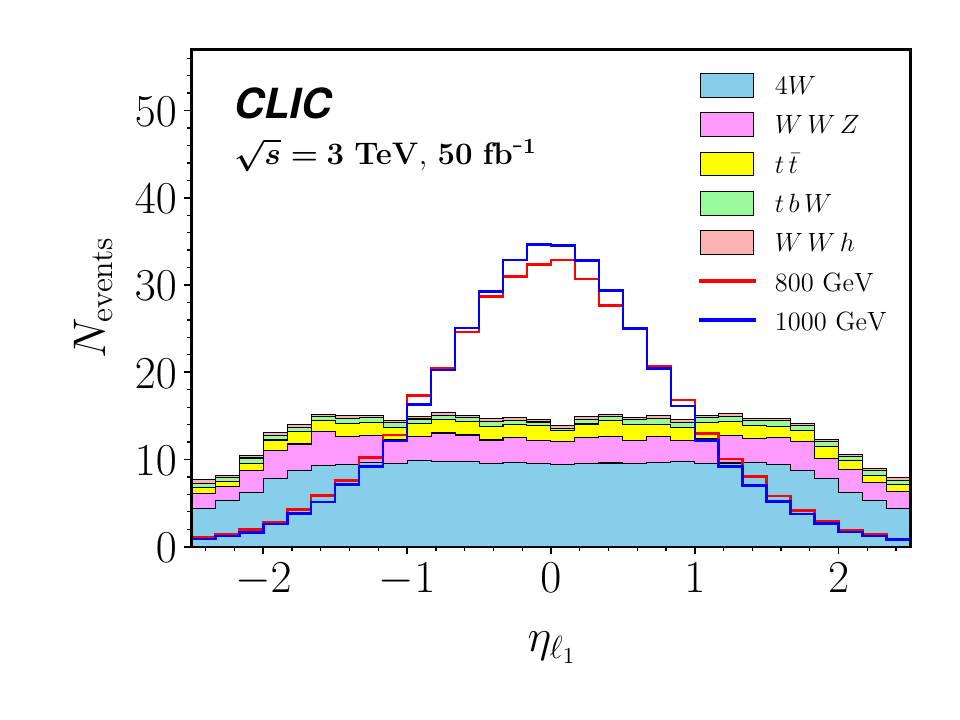}
\end{minipage}
\begin{minipage}{0.45\textwidth}
\includegraphics[width=1.0\textwidth]{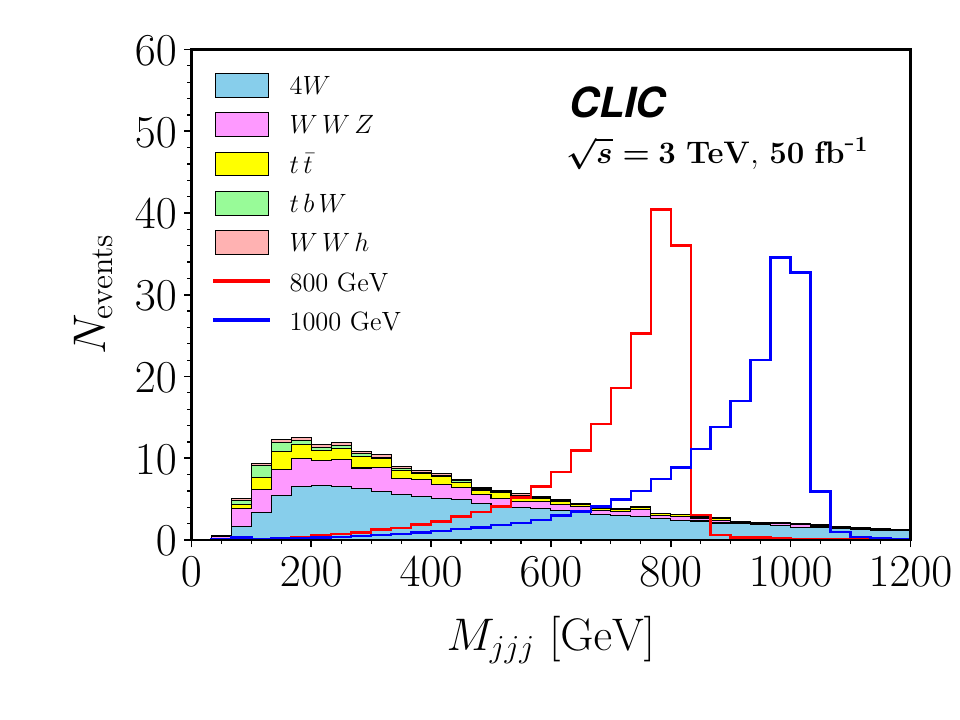}
\end{minipage}
\begin{minipage}{0.45\textwidth}
\includegraphics[width=1.0\textwidth]{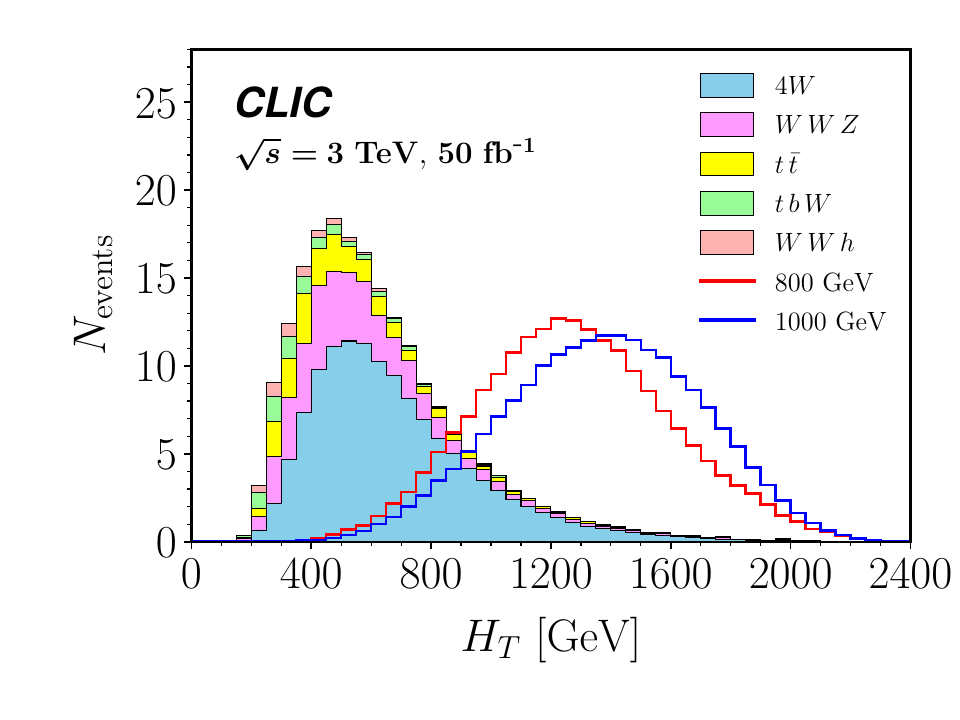}
\end{minipage}
\caption{Kinematic distributions for the signal $\gamma \gamma \rightarrow H^{++}H^{--} \rightarrow 4W \rightarrow \ell^{\pm} \ell^{\pm} + \geq 3j$ and its backgrounds at $\sqrt{s} = 3~ \mathrm{TeV}$ CLIC. The signal yield is normalized to the total background.}
\label{fig3}
\end{figure}

\begin{table}[htbp]
  \centering
  \setlength{\tabcolsep}{7pt}
  \begin{tabular}{!{\vrule width 1.2pt} c | c | c c c c c!{\vrule width 1.2pt}}
    \Hline
    & & baseline  & $p_{T,\, \ell_1}$ & $\eta_{\ell_1}$ & $M_{jjj}$ & $H_T$ \\
    \Hline
    background & $N_{\text{bkg}}$ & $202$ & $79.2$ & $54.9$ & $23.2$ & $13.9$ \\
    \Hline
    \multirow{2}{*}{$m_{H^{\pm\pm}} (800)$} & $N_{\text{signal}}$ & $95.0$ & $81.3$ & $79.2$ & $72.4$ & $70.6$ \\
    & $\mathcal{S}$ & $5.51$ & $6.42$ & $6.84$ & $7.40$ & $7.68$ \\
    \Hline
    \multirow{2}{*}{$m_{H^{\pm\pm}} (1000)$} & $N_{\text{signal}}$ & $23.9$ & $21.6$ & $21.1$ & $20.2$ & $19.8$ \\
    & $\mathcal{S}$ & $1.59$ & $2.15$ & $2.42$ & $3.07$ & $3.41$ \\
    \Hline
  \end{tabular}
  \caption{\label{tab:aa-cutflow}
  Cutflow for the signal $\gamma \gamma \rightarrow H^{++}H^{--} \rightarrow 4W \rightarrow \ell^{\pm} \ell^{\pm} + \geq 3j$ and its background at $\sqrt{s} = 3~ \mathrm{TeV}$ CLIC, along with the corresponding statistical significance.}
\end{table}

\par
The cutflow for signal and background events is presented in Table \ref{tab:aa-cutflow}. Upon applying the optimized selection criteria in Eq.\eqref{eq:optimized-selection}, more than $70\%$ of the signal events are retained, while the total background is reduced to below $10\%$ of its yield after baseline selection. For a doubly charged Higgs boson with a mass of around $1~\mathrm{TeV}$, approximately $20$ signal events are expected at $\sqrt{s} = 3~ \mathrm{TeV}$ CLIC following the optimized event selection, assuming an integrated luminosity of $50~\mathrm{fb}^{-1}$. These results demonstrate that the $\gamma\gamma$ collision mode at CLIC provides a promising experimental platform for searching for doubly charged Higgs bosons below the TeV scale. For low-mass doubly charged Higgs bosons, although the signal significance is already sufficiently high, it can still be improved further by refining the kinematic cuts \eqref{eq:optimized-selection}.

\subsection{$e^+ e^- \rightarrow H^{++} H^{--} \rightarrow W^+ W^+ W^- W^-$ at BP2}
\label{sec:4.4}
\par
Apart from $\gamma\gamma \rightarrow H^{++} H^{--}$ studied in subsection \ref{sec:4.3}, the $e^+e^-$ collision mode provides another major production mechanism for doubly charged Higgs boson pairs at lepton colliders in the gauge-like region, as shown in the top-right panel of Fig.\ref{fig2}. Since $H^{\pm\pm}$ predominantly decays into same-sign $W$ boson pairs in this region, the resulting signal process is $e^+ e^- \rightarrow H^{++} H^{--} \rightarrow W^+ W^+ W^- W^-$. Analogous to the $\gamma\gamma \rightarrow H^{++}H^{--} \rightarrow 4W$ signal process studied in the previous subsection, we focus exclusively on the $\ell^\pm \ell^\pm + \geq 3j$ final state from the $4W$ system to explore the discovery potential of the doubly charged Higgs boson in $e^+e^-$ collisions. Although the $e^+ e^-$ collision mode targets the same final state, its background composition differs substantially from that of the $\gamma\gamma$ mode, primarily due to its fixed center-of-mass energy and the dominance of the $s$-channel production mechanism. Consequently, the dominant backgrounds in the $e^+ e^-$ mode are
\begin{align}
&
(1)
&&
 W^+ W^- \ell^+ \ell^-\,,
 &~~&
 (2)
 &&
 W^+ W^- W^\pm \ell^\mp \nu\,, 
 &~~&
 (3)
 &&
W^\pm Z \ell^\mp \nu\,,
 &~~&
 (4)
 &&
 W^+ W^- Z \ell^+ \ell^-\,,
 \nonumber
 \\
 &
 (5)
 &&
 W^+ W^+ W^- W^-\,,
 &~~&
 (6)
 &&
 W^+ W^- Z\,,
 &~~&
 (7)
 &&
 W^+ W^- Z Z\,.
\end{align}
All these backgrounds arise from prompt production. The SSDL signature typically consists of one non-resonant lepton and one lepton from a vector-boson decay, with contributions from jet misidentification being negligible. The first two backgrounds are dominant, accounting for approximately $85\%$ of the total, whereas the last three, which arise solely from resonant production, contribute only a minor fraction, less than $5\%$.

\par
Table \ref{tab:epem-sgnf} presents the expected numbers of signal and background events, together with the corresponding signal significance, for five representative mass points in the range of $[400, 1200]~\mathrm{GeV}$. The results demonstrate that in the low-mass region, the signal significance markedly exceeds $10\sigma$. It is worth noting that the mass of the doubly charged Higgs boson induces two competing effects. In high-energy collisions, the Lorentz boost of an unstable particle typically leads to the collimation of its decay products, which in turn reduces the efficiency of both object reconstruction and event selection. For the $e^+e^- \rightarrow H^{\pm\pm}H^{\mp\mp} \rightarrow 4W$ process under study at $3~\mathrm{TeV}$ CLIC, an increase in $m_{H^{\pm\pm}}$ results in a more balanced distribution of the energy and directions of the four $W$ bosons. This change weakens the overall impact of the Lorentz boost, thereby mitigating the loss in event reconstruction and selection efficiency. Conversely, the production cross section of this process decreases with increasing $m_{H^{\pm\pm}}$ due to the phase-space suppression of the doubly charged Higgs pair. The interplay between these two effects results in a maximum signal significance at an intermediate value of $m_{H^{\pm\pm}}$ around $600~ \mathrm{GeV}$.

\begin{table}[htbp]
  \centering
  \setlength{\tabcolsep}{7pt}
  \begin{tabular}{!{\vrule width 1.2pt} c | c c | c c c c c!{\vrule width 1.2pt}}
    \Hline
    $\sqrt{s}~[\mathrm{TeV}]$ & \multicolumn{2}{c |}{$1.5$} & \multicolumn{5}{c!{\vrule width 1.2pt}}{$3.0$} \\
    $m_{H^{\pm\pm}}~[\mathrm{GeV}]$ & $400$ & $600$ & $400$ & $600$ & $800$ & $1000$ & $1200$ \\
    \Hline
    $N_{\text{signal}} \,/\, 10^3$ & $2.57$ & $1.04$ & $0.954$ & $1.05$ & $0.880$ & $0.588$ & $0.258$ \\
    $N_{\text{bkg}} \,/\, 10^3$ & \multicolumn{2}{c |}{$1.49$} & \multicolumn{5}{c!{\vrule width 1.2pt}}{$4.32$} \\
    $\mathcal{S}$ & $40.3$ & $20.7$ & $13.1$ & $14.3$ & $12.2$ & $8.39$ & $3.81$ \\
    \Hline
  \end{tabular}
  \caption{\label{tab:epem-sgnf}
  Same as Table \ref{tab:aa-sgnf}, but for $e^+e^- \rightarrow H^{++}H^{--} \rightarrow 4W \rightarrow \ell^\pm \ell^\pm + \geq 3j$.}
\end{table}

\par
To enhance the modest signal significance near the pair production threshold at $3~\mathrm{TeV}$ CLIC, a set of cuts on kinematic variables is introduced to improve the signal-background discrimination. These variables include
\begin{align}
&
(1)
&~&
\slashed{p}_T:
&&
\text{the missing transverse momentum;}
\nonumber
\\
&
(2)
&~&
\Delta R(\ell\,, \ell):
&&
\text{the angular separation between the same-sign leptons;}
\nonumber
\\
&
(3)
&~&
\Delta \phi(\ell\ell\,, \slashed{p}_T):
&&
\text{the azimuthal angle difference between the SSDL system and $\slashed{p}_T$;}
\nonumber
\\
&
(4)
&~&
H_T:
&&
\text{the scalar sum of the jet transverse momenta and $\slashed{p}_T$.}
\nonumber
\end{align}
Their distributions for signal and background processes are depicted in Fig.\ref{fig4}, from which we define the following event selection criteria:
\begin{enumerate}
\item Signal events are characterized by a larger $\slashed{p}_T$, and thus we require $\slashed{p}_T > 150~\mathrm{GeV}$;
\item The same-sign lepton pair from $H^{\pm\pm}$ decays is typically more collimated, which motivates the requirement $\Delta R(\ell,\, \ell) < 3$;
\item In signal events, the transverse momentum of the SSDL system typically aligns with $\slashed{p}_T$, thereby motivating the requirement $\big|\Delta \phi(\ell\ell,\, \slashed{p}_T)\big| < 1.5$;
\item The jets in signal events originate from the decays of the doubly charged Higgs and typically yield larger values of $H_T$, for which we impose the requirement $H_T > 1500~\mathrm{GeV}$.
\end{enumerate}

\begin{figure}[htbp]
\centering
\begin{minipage}{0.45\textwidth}
\includegraphics[width=1.0\textwidth]{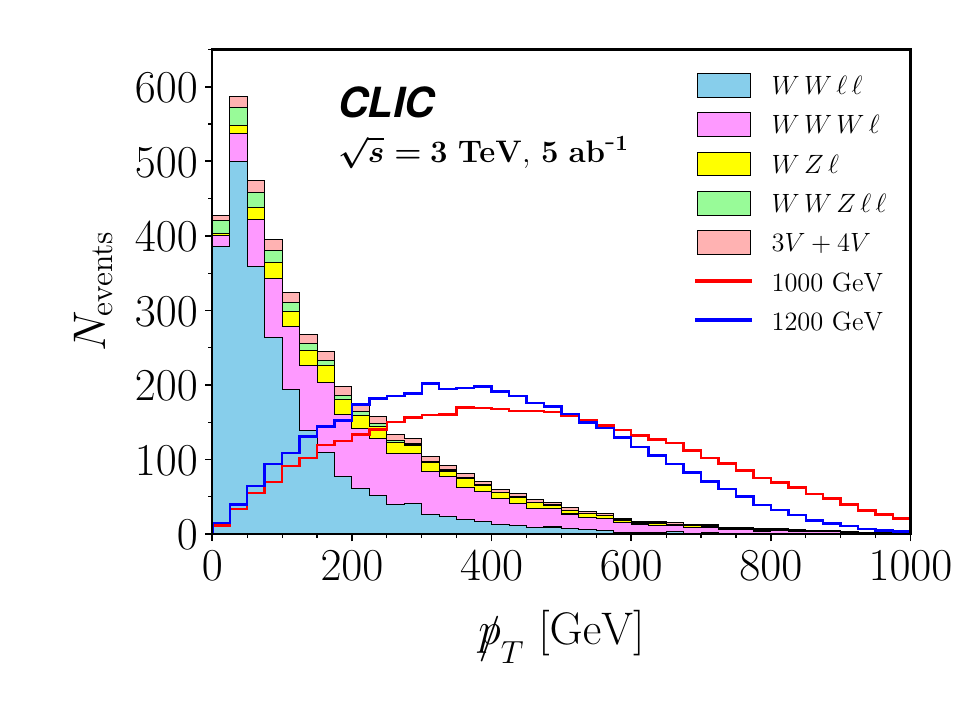}
\end{minipage}
\begin{minipage}{0.45\textwidth}
\includegraphics[width=1.0\textwidth]{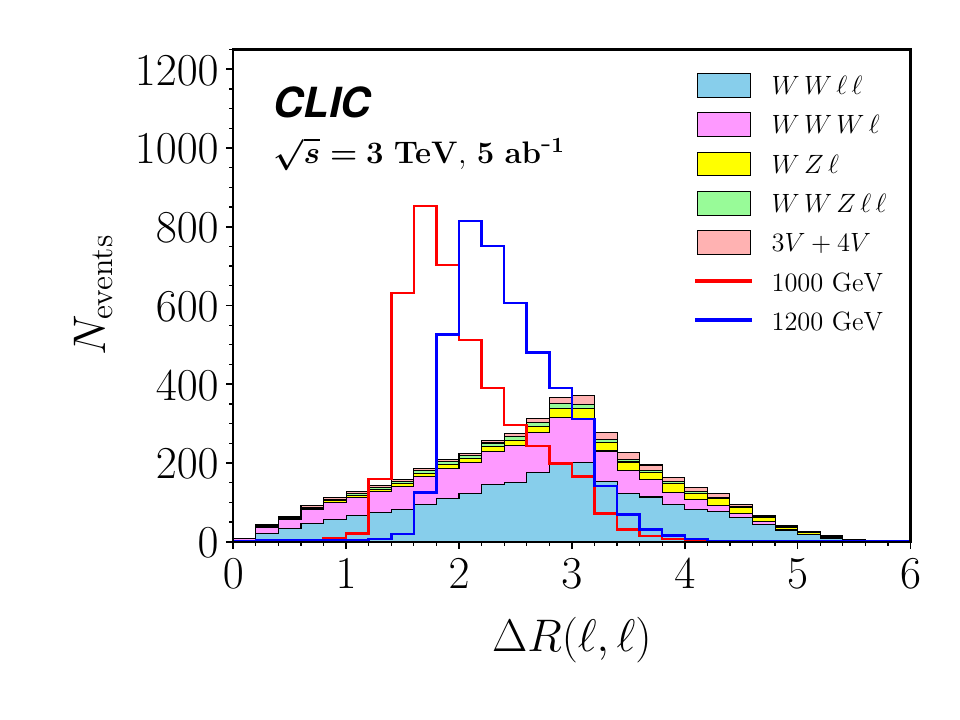}
\end{minipage}
\begin{minipage}{0.45\textwidth}
\includegraphics[width=1.0\textwidth]{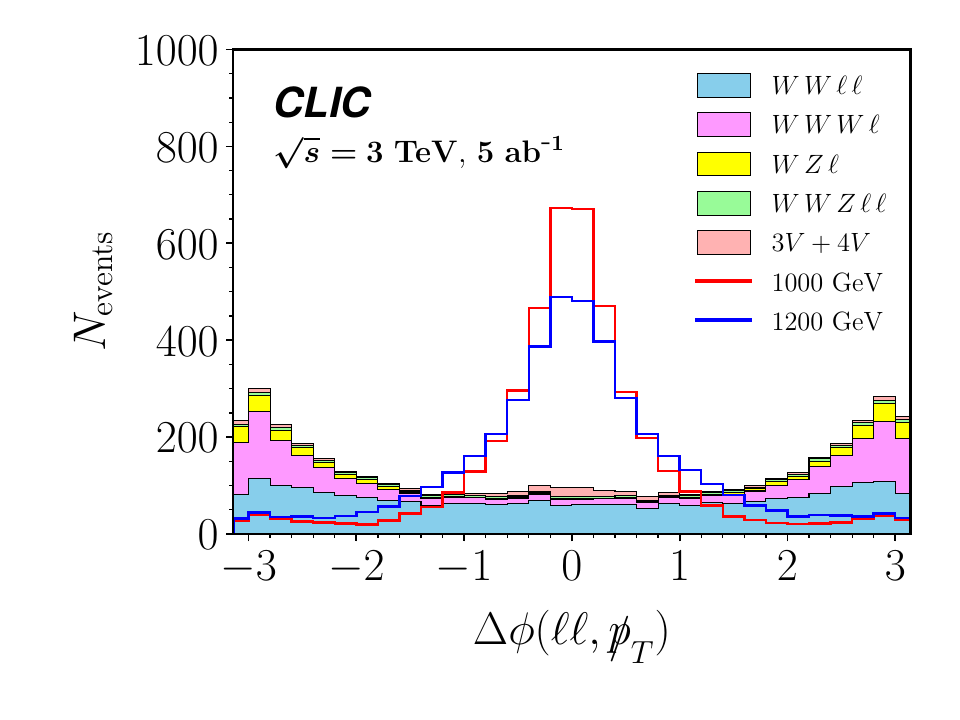}
\end{minipage}
\begin{minipage}{0.45\textwidth}
\includegraphics[width=1.0\textwidth]{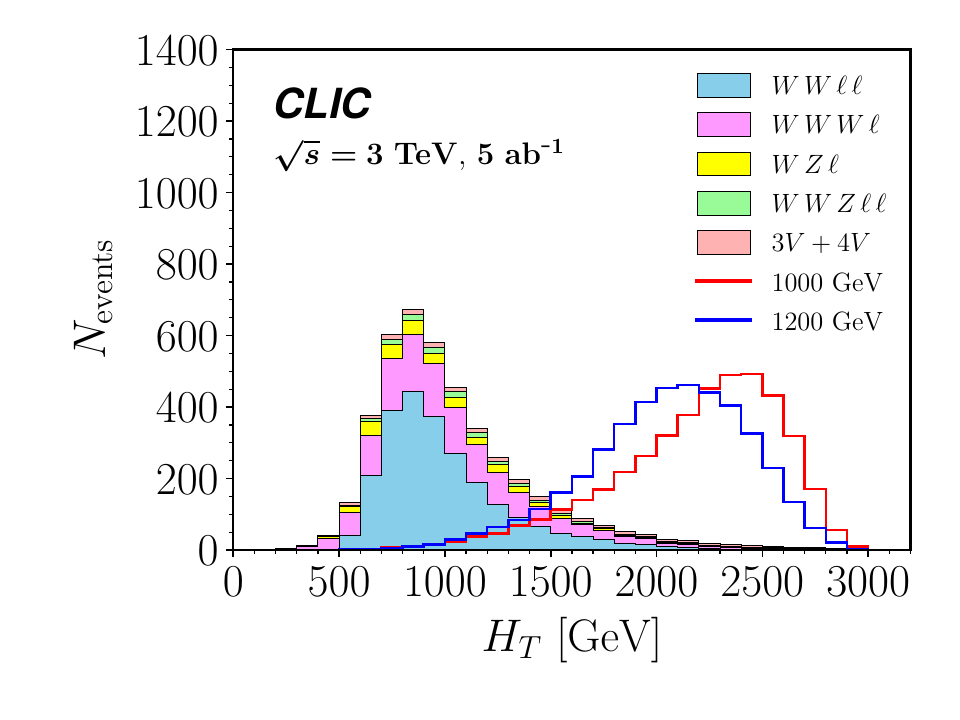}
\end{minipage}
\caption{Same as Fig.\ref{fig3}, but for $e^+e^- \rightarrow H^{++}H^{--} \rightarrow 4W \rightarrow \ell^{\pm} \ell^{\pm} + \geq 3j$.}
\label{fig4}
\end{figure}

\par
Table \ref{tab:epem-cutflow} details the cutflow of signal and background yields, along with the resulting significance. The requirements on $\Delta\phi(\ell\ell,\, \slashed{p}_T)$ and $H_T$ lead to the most significant improvement in signal significance, consistent with the distributions shown in Fig.\ref{fig4}. The final results indicate that, even for $m_{H^{\pm\pm}} = 1200~\mathrm{GeV}$, the signal significance exceeds $10\sigma$. This demonstrates that the $e^+ e^-$ mode offers strong sensitivity to the doubly charged Higgs boson, covering nearly the entire kinematically accessible mass range. Compared to the $\gamma \gamma$ collision, the $e^+ e^-$ mode produces roughly an order of magnitude more signal events at $m_{H^{\pm\pm}} = 1000~\mathrm{GeV}$, primarily due to higher integrated luminosity. At the same integrated luminosity, however, the $\gamma \gamma$ mode exhibits superior performance. For instance, at an integrated luminosity of $50~\mathrm{fb}^{-1}$ for $m_{H^{\pm \pm}} = 1000~\mathrm{GeV}$, the $e^+ e^-$ mode retains only about $4$ signal events, while the $\gamma \gamma$ mode produces approximately $19$, after applying the optimized event selection criteria.

\begin{table}[htbp]
  \centering
  \setlength{\tabcolsep}{7pt}
  \begin{tabular}{!{\vrule width 1.2pt} c | c | c c c c c!{\vrule width 1.2pt}}
    \Hline
    & & baseline  & $\slashed{p}_T$ & $\Delta R (\ell, \ell)$ & $\Delta \phi (\ell\ell, \slashed{p}_T)$ & $H_T$ \\
    \Hline
    background & $N_{\text{bkg}}$ & $4320$ & $1850$ & $1060$ & $216$ & $64.2$ \\
    \Hline
    \multirow{2}{*}{$m_{H^{\pm\pm}} (1000)$} & $N_{\text{signal}}$ & $588$ & $539$ & $504$ & $461$ & $448$ \\
    & $\mathcal{S}$ & $8.39$ & $11.0$ & $12.7$ & $17.7$ & $19.8$ \\
    \Hline
    \multirow{2}{*}{$m_{H^{\pm\pm}} (1200)$} & $N_{\text{signal}}$ & $258$ & $231$ & $202$ & $172$ & $165$ \\
    & $\mathcal{S}$ & $3.81$ & $5.06$ & $5.69$ & $8.73$ & $10.9$ \\
    \Hline
  \end{tabular}
  \caption{\label{tab:epem-cutflow}
  Same as Table \ref{tab:aa-cutflow}, but for $e^+e^- \rightarrow H^{++}H^{--} \rightarrow 4W \rightarrow \ell^{\pm} \ell^{\pm} + \geq 3j$.}
\end{table}

\section{Discovery potential at 14 TeV HL-LHC}
\label{sec:5}
\par
For comparison, we explore the discovery potential of the doubly charged Higgs boson at the 14 TeV HL-LHC, with an integrated luminosity of $3~\mathrm{ab}^{-1}$. The signal process under consideration is Drell-Yan pair production, $pp \rightarrow H^{\pm\pm} H^{\mp\mp}$. The analysis is conducted separately in the Yukawa-like and gauge-like regions. Both signal and background processes are simulated using the same computational framework developed for the CLIC study. Events are generated at the matrix-element level with up to two partons and subsequently matched to parton shower using \texttt{Pythia8}. To mitigate the more challenging background at hadron colliders, a tighter lepton isolation criterion is imposed: the scalar sum of the transverse momenta of all particles within a cone of radius $R = 0.5$ around the lepton, excluding the lepton itself, is required to be less than $12\%$ of the lepton transverse momentum. Jets in the final state are reconstructed using the anti-$k_t$ algorithm with a radius parameter of $R = 0.4$. The baseline selection applied to final-state objects is identical to that used in the CLIC analysis, as defined in Eq.\eqref{eq:bs}.

\subsection{$p p \rightarrow H^{++} H^{--} \rightarrow e^+ e^+ e^- e^-$ at BP1}
\label{sec:5.1}
\par
In the Yukawa-like region, the $H^{\pm\pm}$ boson predominantly decays into same-sign lepton pairs. Under the single-dominance hypothesis, the signal process we focus on is $pp \rightarrow H^{++} H^{--} \rightarrow e^+ e^+ e^- e^-$. At BP1, the pair-production cross section of $H^{\pm\pm}$ at the $14~\mathrm{TeV}$ LHC is approximately four orders of magnitude smaller, or potentially even more, depending on the doubly charged Higgs mass, compared to the dominant production processes at CLIC. Nonetheless, owing to the clean four-lepton final state and the high integrated luminosity, a substantial signal significance remains expected.

\par
At hadron colliders, backgrounds are typically much larger than at lepton colliders. By requiring a final state with four electrons that satisfy the baseline selection criteria, the major backgrounds can be classified into three distinct categories:
\begin{itemize}
\item \textbf{Prompt lepton backgrounds}: including $ZZ$, $Z\ell^+\ell^-$, $\ell^+\ell^-\ell^+\ell^-$, $VVZ~ (V = W,\, Z)$, and $t\bar{t}Z$. In all cases, the vector bosons decay leptonically into electrons or $\tau$-leptons, with the $\tau$-leptons undergoing cascade decays to electrons.
\item \textbf{Fake lepton backgrounds}: originating from $WV+$jets and $t \bar{t}$ events, where at least one electron is misidentified, such as from jets or $b$-hadrons.
\item \textbf{$\gamma$-conversion backgrounds}: for example, the Drell-Yan process $pp \rightarrow Z/\gamma^{\ast} \rightarrow \ell^+ \ell^-$. An additional electron pair is produced by the conversion of an extra photon radiated off either the initial or final states.
\end{itemize}
Among these backgrounds, $\gamma$ conversion constitutes the dominant contribution, making up approximately $80\%$ of the total. Resonant $ZZ$ production is the subleading background, contributing about $10\%$, while all others remain relatively minor. 

\par
After applying the baseline event selection criteria, the background remains overwhelmingly dominant over the signal. Due to the relatively large mass of the doubly charged Higgs boson in the Yukawa-like scenario, the signal and background exhibit clearly distinct invariant mass distributions for both same-sign and opposite-sign electron pairs, as illustrated in Fig.\ref{fig5}. For background events, the invariant mass of the leading opposite-sign electron pair $M_{e_1^+ e_1^-}$, along with that of the same-sign electron pairs, such as $M_{e^+e^+}$, is concentrated in the low-mass region. In contrast, for signal events, $M_{e^+_1 e^-_1}$ predominantly occupies the higher-mass region, while $M_{e^+ e^+}$ exhibits a distinct peak near the doubly charged Higgs mass. Therefore, we introduce the following additional selection criteria on top of the baseline requirements:
\begin{equation}
M_{e^+_1 e^-_1} > 500~\mathrm{GeV}\,,
\qquad\qquad
M_{e^\pm e^\pm} > 800~\mathrm{GeV}\,,
\end{equation}
which effectively suppresses the backgrounds. Table \ref{tab:pp1-cutflow} summarizes the signal and background event yields after each stage of the cutflow (baseline, $M_{e_1^+e_1^-}$, and $M_{e^{\pm}e^{\pm}}$), together with the corresponding statistical significance. At the HL-LHC with an integrated luminosity of $3~\mathrm{ab}^{-1}$, a doubly charged Higgs boson with a mass of $1.1~ \mathrm{TeV}$ is expected to yield approximately $19$ signal events, corresponding to a discovery potential slightly exceeding $4\sigma$. In the Yukawa-like region, the small production cross section of the signal limits the HL-LHC's sensitivity to a relatively narrow mass window for the doubly charged Higgs boson. Therefore, the CLIC offers superior discovery potential compared to the HL-LHC for doubly charged Higgs bosons in the Yukawa-like scenario.

\begin{figure}[htbp]
\centering
\begin{minipage}{0.45\textwidth}
\includegraphics[width=1.0\textwidth]{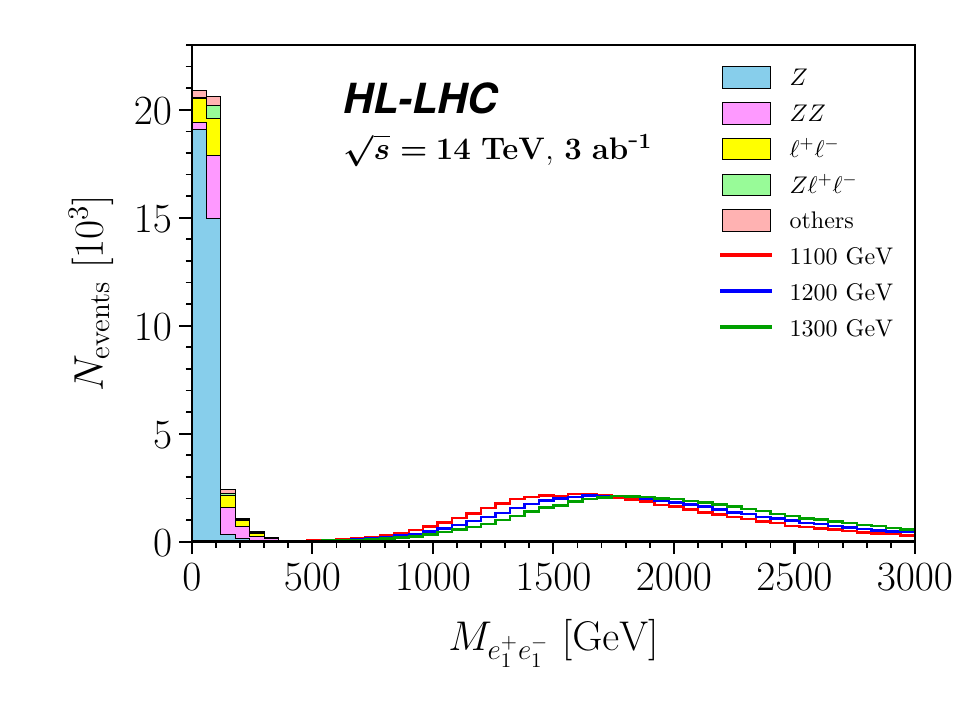}
\end{minipage}
\begin{minipage}{0.45\textwidth}
\includegraphics[width=1.0\textwidth]{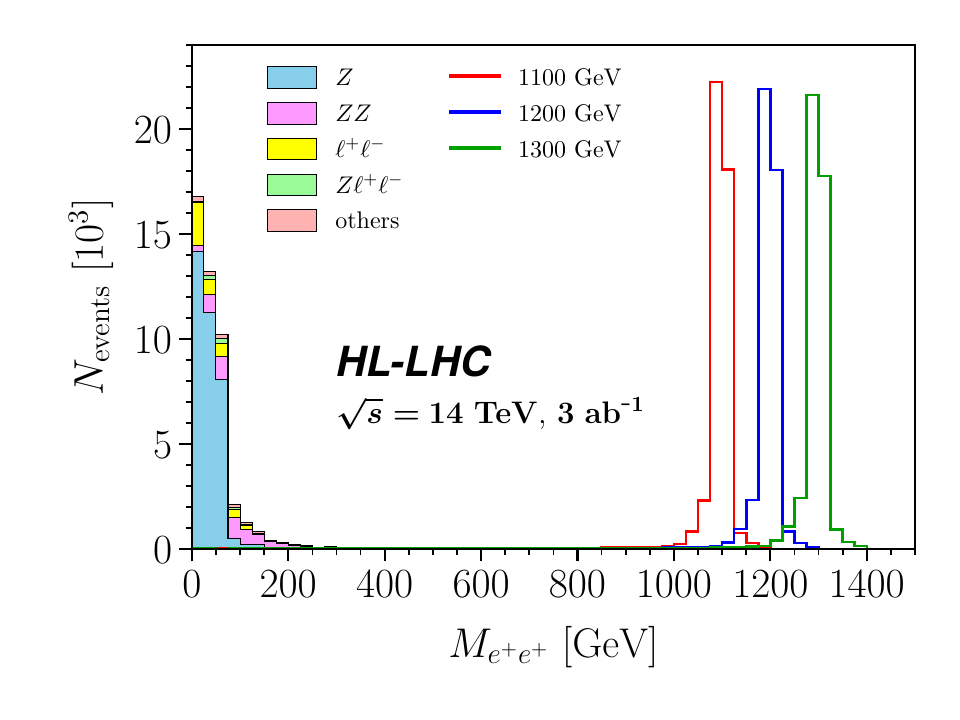}
\end{minipage}
\caption{Kinematic distributions for the signal process $pp \rightarrow H^{++} H^{--} \rightarrow e^+ e^+ e^- e^-$ and its backgrounds at the $14~\mathrm{TeV}$ HL-LHC. The signal yield is normalized to the total background.}
\label{fig5}
\end{figure}

\begin{table}[htbp]
  \centering
  \setlength{\tabcolsep}{7pt}
  \begin{tabular}{!{\vrule width 1.2pt} c | c | c c c!{\vrule width 1.2pt}}
    \Hline
    & & baseline  & $M_{e^+_1 e^-_1}$ & $M_{e^\pm e^\pm}$ \\[3pt]
    \Hline
    background & $N_{\text{bkg}}$ & $45800$ & $70.2$ & $1.32$ \\
    \Hline
    \multirow{2}{*}{$m_{H^{\pm\pm}} (1100)$} & $N_{\text{signal}}$ & $19.5$ & $19.5$ & $19.1$ \\
    & $\mathcal{S}$ & $0.0913$ & $2.06$ & $4.23$ \\
    \Hline
    \multirow{2}{*}{$m_{H^{\pm\pm}} (1200)$} & $N_{\text{signal}}$ & $11.0$ & $11.0$ & $10.8$ \\
    & $\mathcal{S}$ & $0.0516$ & $1.22$ & $3.11$ \\
    \Hline
    \multirow{2}{*}{$m_{H^{\pm\pm}} (1300)$} & $N_{\text{signal}}$ & $6.31$ & $6.30$ & $6.21$ \\
    & $\mathcal{S}$ & $0.0295$ & $0.720$ & $2.26$ \\
    \Hline
  \end{tabular}
  \caption{\label{tab:pp1-cutflow}
  Cutflow for the signal process $pp \rightarrow H^{++} H^{--} \rightarrow e^+ e^+ e^- e^-$ and its background at the $14~ \mathrm{TeV}$ HL-LHC, along with the corresponding significance.}
\end{table}

\subsection{$p p \rightarrow H^{++} H^{--} \rightarrow W^+ W^+ W^- W^-$ at BP2}
\label{sec:5.2}
\par
In the gauge-like region, $H^{\pm\pm}$ predominantly decays into same-sign $W$-boson pairs. In this subsection, we investigate the process $p p \rightarrow H^{++} H^{--} \rightarrow W^+ W^+ W^- W^-$ at BP2 to evaluate the discovery potential of the doubly charged Higgs boson via its bosonic decay at the $14~\mathrm{TeV}$ HL-LHC. Following the event selection strategy of the CLIC $4W$ analysis, we target the final state with a same-sign lepton pair and at least three jets (excluding $b$-jets). The signal events primarily arise from $4W \rightarrow \ell^{\pm} \ell^{\pm} + 2\nu + 4q$, with a smaller contribution from $4W \rightarrow \ell^+ \ell^- \ell^{\pm} + 3\nu + 2q$. Accordingly, the dominant backgrounds fall into two categories:
\begin{itemize}
\item \textbf{Prompt lepton backgrounds}: including multiboson, $t\bar{t}V$, and $Vh$ ($h \rightarrow VV^{\ast}$). In this category, SSDLs originate from the leptonic decays of vector bosons, with $WZ \rightarrow \ell \nu \ell^+ \ell^-$ being the dominant contribution. $ZZ$, $t\bar{t}W$, and same-sign $W$-boson pair also contribute noticeably. Other processes, including triboson and $Vh$ production, contribute only marginally due to their relatively small production cross sections and the tendency of their final states to involve more than two leptons, which leads to efficient rejection by the baseline selection.

 \item \textbf{Fake lepton backgrounds}: originating from $V+$jets and $t\bar{t}$ events.
 
In these backgrounds, at least one lepton originates from jet-to-lepton misidentification. Although the misidentification rate is small, the very large production cross sections of these processes result in more background events than the prompt production. Of these, $W+$jets is the dominant contributor, followed by $t\bar{t}$, with $Z+$jets making a comparatively smaller contribution.
\end{itemize}
In addition to the aforementioned backgrounds, events with charge misidentification may also contribute as a potential background. Studies reported in Ref.\cite{ATLAS:2021jol} indicate that, after the final event selection, the contribution from this type of backgrounds is negligible. Moreover, with the exceptional tracking performance anticipated at the HL-LHC, the charge misidentification rate can be reduced to an exceedingly low level. As a result, these backgrounds are not considered in the current analysis.

\par
After the baseline event selection, background events dominate overwhelmingly, surpassing the signal yield by three to four orders of magnitude. We therefore further compare several characteristic kinematic distributions of the signal and background processes, exploiting their differences to achieve more efficient background suppression and, consequently, enhance the signal significance. As shown in Fig.\ref{fig6}, these discriminating variables include the invariant masses of the final-state lepton pair and the three leading jets, $M_{\ell\ell}$ and $M_{jjj}$, as well as $\slashed{p}_T$, $\Delta\phi(\ell\ell,\, \slashed{p}_T)$ and $H_T$. Due to the large mass of the doubly charged Higgs boson, kinematic variables related to energy and momentum, such as $\slashed{p}_T$, $M_{\ell\ell}$, $H_T$, and $M_{jjj}$, take on higher values in signal events. Both $H_T$ and $M_{jjj}$ distributions exhibit a strong dependence on $m_{H^{\pm\pm}}$: $H_T$ features a pronounced peak around $2\, m_{H^{\pm\pm}}$, while $M_{jjj}$ shows a peak that decreases sharply near $m_{H^{\pm\pm}}$, indicating that these leading jets predominantly originate from the decay of $H^{\pm\pm}$. Motivated by these kinematic features, we implement the following optimized selection criteria to improve the signal significance:
\begin{equation}
\label{eq:lhcbs}
\begin{aligned}
&
\slashed{p}_T > 120~\mathrm{GeV} \,,
\qquad~~
M_{\ell\ell} > 80~\mathrm{GeV}\,,
\qquad~~
H_T > 2\, m_{H^{\pm\pm}}\,,
\\
&
3/4\, m_{H^{\pm\pm}} < M_{jjj} < m_{H^{\pm\pm}} + 50~\mathrm{GeV}\,.
  \end{aligned}
\end{equation}
Here, the upper bound of the $M_{jjj}$ selection window includes an additional $50~\text{GeV}$ margin to compensate for the broadening of the signal peak caused by the finite jet energy resolution. Despite the significant difference in the $\Delta\phi(\ell\ell,\, \slashed{p}_T)$ distributions between signal and background, no cut is imposed on $\Delta\phi(\ell\ell,\, \slashed{p}_T)$ due to its correlation with $\slashed{p}_T$ and $M_{\ell\ell}$. Once cuts on $\slashed{p}_T$ and $M_{\ell\ell}$ are applied, its discriminatory power is largely redundant.

\begin{figure}[htbp]
\centering
\begin{minipage}{0.45\textwidth}
\includegraphics[width=1.0\textwidth]{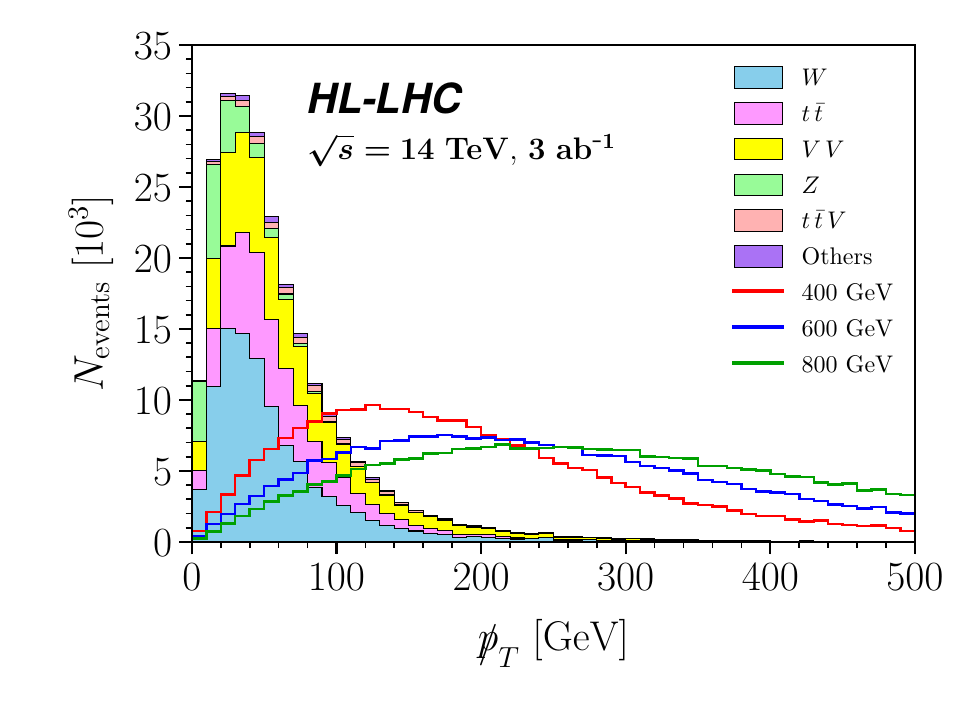}
\end{minipage}
\begin{minipage}{0.45\textwidth}
\includegraphics[width=1.0\textwidth]{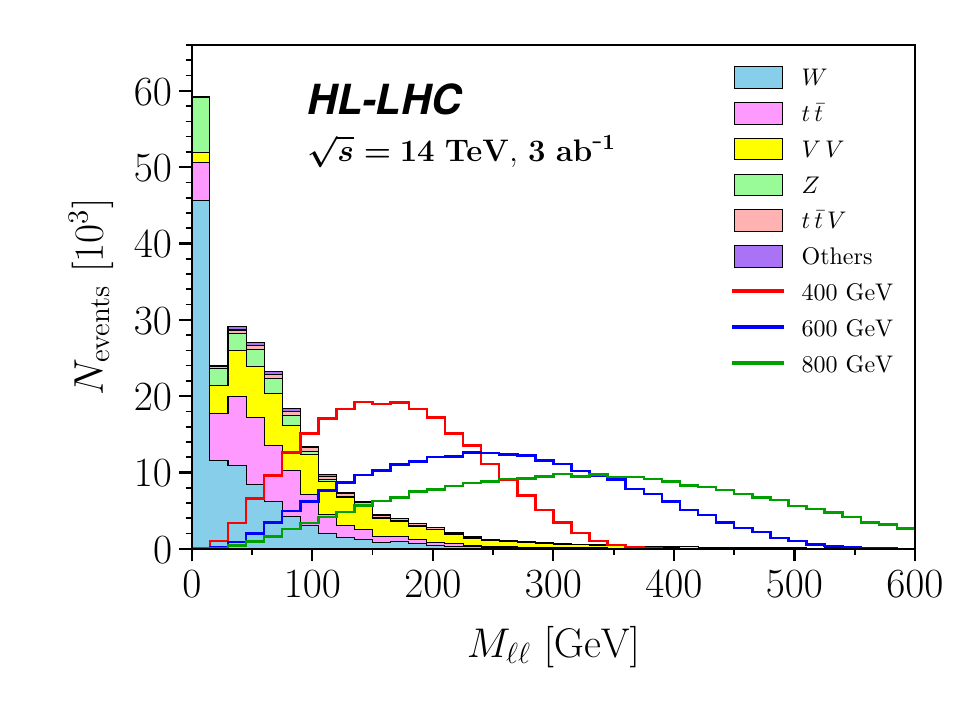}
\end{minipage}
\begin{minipage}{0.45\textwidth}
\includegraphics[width=1.0\textwidth]{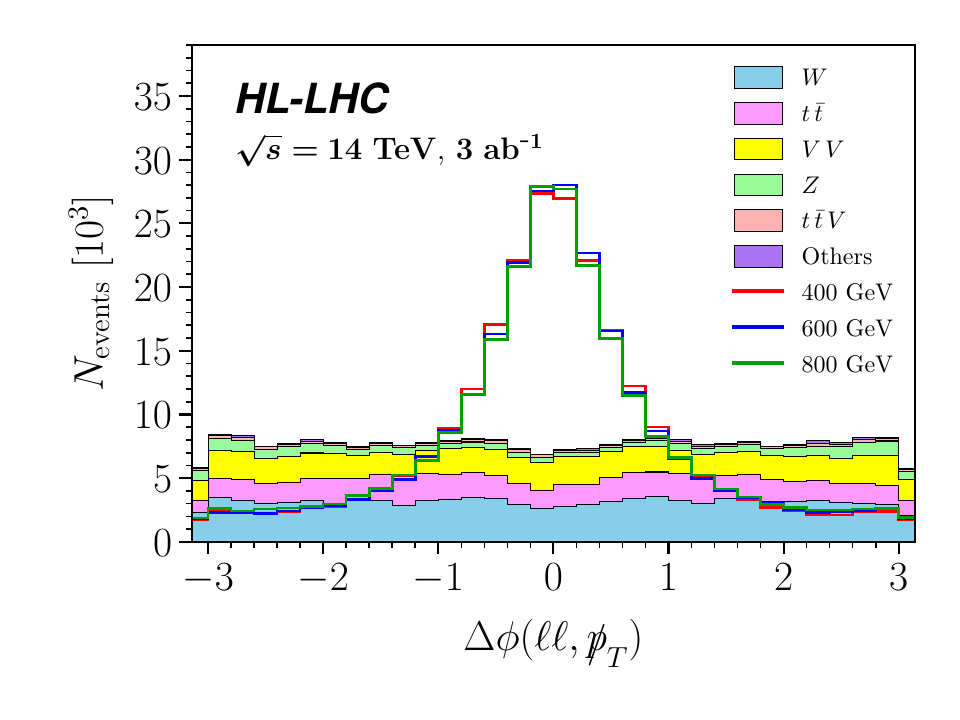}
\end{minipage}
\begin{minipage}{0.45\textwidth}
\includegraphics[width=1.0\textwidth]{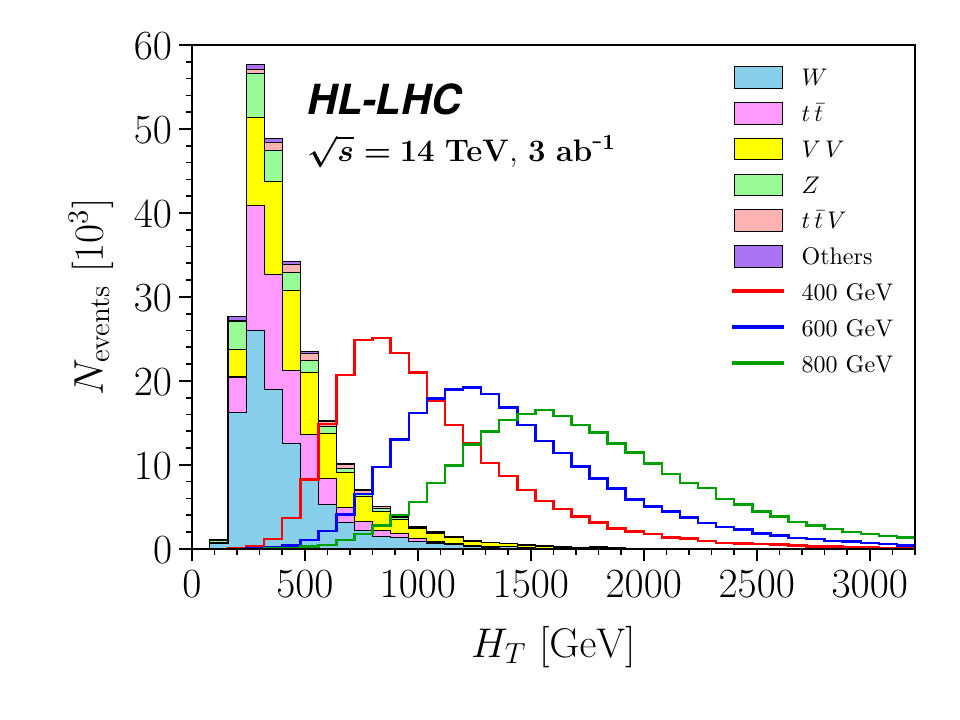}
\end{minipage}
\begin{minipage}{0.45\textwidth}
\includegraphics[width=1.0\textwidth]{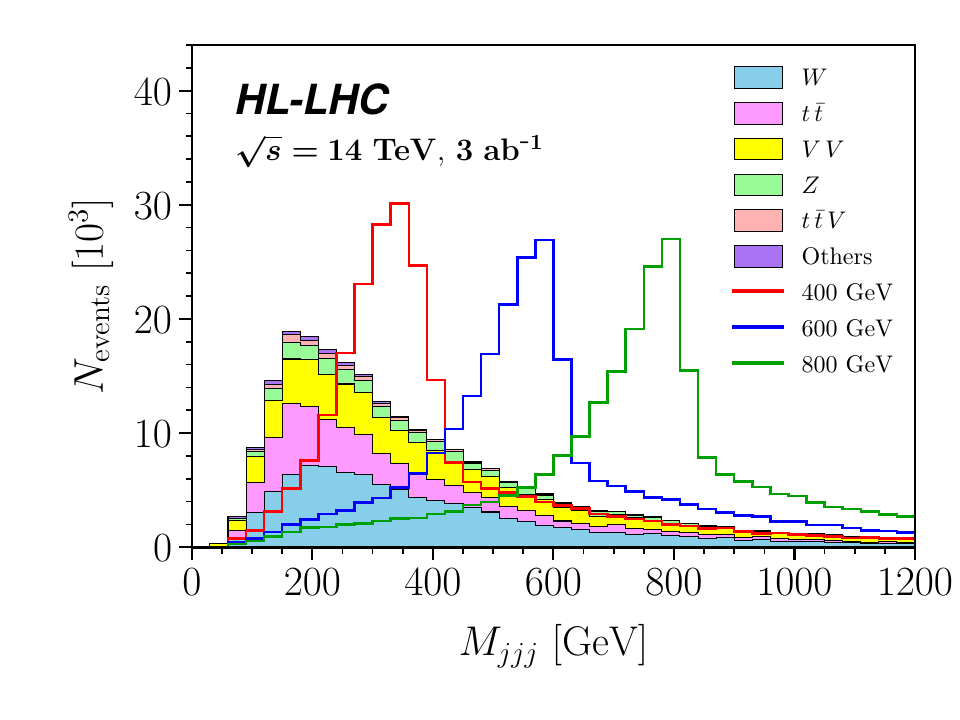}
\end{minipage}
\caption{Same as Fig.\ref{fig5}, but for $p p \rightarrow H^{++} H^{--} \rightarrow 4W \rightarrow \ell^\pm \ell^\pm + \geq 3j$.}
\label{fig6}
\end{figure}

\par
The cutflow of signal and background yields, along with the corresponding significance, is summarized in Table \ref{tab:pp2-cutflow}. It is evident that the selection efficiency for both signal and background events demonstrates a distinctly different dependence on the doubly charged Higgs mass. For the signal, the efficiency remains relatively stable across $m_{H^{\pm\pm}}$, roughly $1/4$. However, for the background, the selection efficiency decreases with increasing $m_{H^{\pm\pm}}$. For instance, as $m_{H^{\pm\pm}}$ increases from $400~ \mathrm{GeV}$ to $600~ \mathrm{GeV}$, the efficiency of the event selection criteria in Eq.\eqref{eq:lhcbs} for the background decreases from approximately $3\text{\textperthousand}$ to below $1\text{\textperthousand}$. Assuming an integrated luminosity of $3~\mathrm{ab}^{-1}$, the expected statistical significance is approximately $3\sigma$ at $m_{H^{\pm\pm}} = 400~\mathrm{GeV}$, and decreases significantly as $m_{H^{\pm\pm}}$ increases. This behavior is primarily driven by the rapidly decreasing production cross section with increasing $m_{H^{\pm\pm}}$, as shown in the bottom-right panel of Fig.\ref{fig2}. In comparison, the $3~\mathrm{TeV}$ CLIC offers significant discovery potential for a TeV-scale doubly charged Higgs boson in the gauge-like scenario.

\begin{table}[htbp]
  \centering
  \setlength{\tabcolsep}{6pt}
  \begin{tabular}{!{\vrule width 1.2pt} c | c | c c c c c!{\vrule width 1.2pt}}
    \Hline
    & & baseline  & $\slashed{p}_T$ & $M_{\ell\ell}$ & $H_T$ & $M_{jjj}$ \\
    \Hline
    \multirow{3}{*}{$m_{H^{\pm\pm}} (400)$} & $N_{\text{signal}}$ & $399$ & $283$ & $247$ & $200$ & $90.2$ \\
    & $N_{\text{bkg}}$ & $2.46 \times 10^{5}$ & $2.42 \times 10^{4}$ & $9.09 \times 10^{3}$ & $4.13 \times 10^{3}$ & $647$ \\
    & $\mathcal{S}$ & $0.804$ & $1.81$ & $2.56$ & $3.04$ & $3.32$ \\
    \Hline
    \multirow{3}{*}{$m_{H^{\pm\pm}} (600)$} & $N_{\text{signal}}$ & $68.0$ & $56.4$ & $54.2$ & $37.6$ & $18.7$ \\
    & $N_{\text{bkg}}$ & $2.46 \times 10^{5}$ & $2.42 \times 10^{4}$ & $9.09 \times 10^{3}$ & $1.30 \times 10^{3}$ & $180$ \\
    & $\mathcal{S}$ & $0.137$ & $0.362$ & $0.567$ & $1.03$ & $1.33$ \\
    \Hline
    \multirow{3}{*}{$m_{H^{\pm\pm}} (800)$} & $N_{\text{signal}}$ & $14.6$ & $13.0$ & $12.8$ & $7.74$ & $3.62$ \\
    & $N_{\text{bkg}}$ & $2.46 \times 10^{5}$ & $2.42 \times 10^{4}$ & $9.09 \times 10^{3}$ & $431$ & $65.3$ \\
    & $\mathcal{S}$ & $0.0294$ & $0.0835$ & $0.134$ & $0.370$ & $0.436$ \\
    \Hline
  \end{tabular}
  \caption{\label{tab:pp2-cutflow}
  Same as Table \ref{tab:pp1-cutflow}, but for $p p \rightarrow H^{++} H^{--} \rightarrow 4W \rightarrow \ell^{\pm} \ell^{\pm} + \geq 3j$.}
\end{table}

\section{Summary}
\label{sec:6}
\par
In this work, we detail a comprehensive analysis of the discovery potential for the doubly charged Higgs boson at CLIC, considering $e^-e^-$, $e^-\gamma$, $\gamma\gamma$ and $e^+e^-$ collision modes, within the Higgs triplet model. In the Yukawa-like region, the dominant production mechanism for the doubly charged Higgs boson at CLIC is single production via $e^-e^-$ and $e^-\gamma$ collisions, followed by decay into a same-sign lepton pair. These two production modes can achieve a $5\sigma$ discovery sensitivity for a TeV-scale doubly charged Higgs boson, requiring only $Y_{ee} \sim 0.05\; {\textendash}\;0.15$ under the single-dominance hypothesis, well below the current experimental limit of $Y_{ee} < 0.35$. In contrast, in the gauge-like region, pair production through $\gamma\gamma$ and $e^+e^-$ collisions dominates, with subsequent decay into a same-sign $W$ boson pair. Based on the $\ell^{\pm} \ell^{\pm} + \geq 3j$ event selection and assuming an integrated luminosity of $50~ \mathrm{fb}^{-1}$, the $\gamma\gamma$ mode can achieve a $5\sigma$ discovery significance for $H^{\pm\pm}$ with $m_{H^{\pm\pm}} \lesssim 800~ \mathrm{GeV}$. In the $e^+e^-$ mode, the production cross section of the doubly charged Higgs pair is significantly lower than in the $\gamma\gamma$ mode over most of the kinematically allowed mass range. However, as the primary operational mode of CLIC, $e^+e^-$ collisions accumulate a significantly higher integrated luminosity, enabling a $5\sigma$ discovery reach for $H^{\pm\pm}$ up to a mass of approximately $1200~\mathrm{GeV}$ with an integrated luminosity of $5~\mathrm{ab}^{-1}$.

\par
For comparison, we also evaluate the feasibility of probing the doubly charged Higgs boson in the pair production channel at the HL-LHC. Our analysis shows that the HL-LHC exhibits lower observation significance, primarily due to its substantially smaller production cross section compared to CLIC, compounded by more complex background environments. In both the Yukawa-like and gauge-like regions, the HL-LHC reaches $3\sigma$ observation significance only within a narrow range above the experimental exclusion limit on $m_{H^{\pm\pm}}$, highlighting CLIC's superior discovery potential for the doubly charged Higgs boson.


\vskip 8mm
\noindent{\large\bf Acknowledgments:}
\par
This work is supported by the National Natural Science Foundation of China (Grant No. 12061141005) and the CAS Center for Excellence in Particle Physics (CCEPP).

\bibliographystyle{apsrev4-2}
\bibliography{refs}

\end{document}